\documentclass[12pt,a4paper]{article}
\pdfoutput=1
\usepackage{jheppub}
\usepackage{graphicx}
\usepackage{amssymb}
\usepackage{amsmath}
\usepackage{topcapt}
\usepackage{epstopdf}
\usepackage{subfigure}
\DeclareMathSymbol{\IR}{\mathbin}{AMSb}{"52}

\title{Critical point in a holographic defect field theory}

\author[a,b]{Veselin G. Filev}
\author[c, d]{R. C. Rashkov}

\affiliation[a]{
Institute of Mathematics and Informatics, 
Bulgarian Academy of Sciences,\\ Acad. G. Bonchev Str.,\\
1113 Sofia, Bulgaria.}
\emailAdd{vfilev@math.bas.bg}

\affiliation[b]{School of Theoretical Physics, Dublin Institute for Advanced Studies\\
10 Burlington Road, Dublin 4, Ireland.}
\emailAdd{vfilev@stp.dias.ie}

\affiliation[c]{Department of Physics, Sofia University,\\
5 J. Bourchier Blvd., 1164 Sofia, Bulgaria}

\affiliation[d]{Institute for Theoretical Physics, Vienna University of Technology, \\Wiedner Hauptstr. 8-10, 1040 Vienna, Austria}

\abstract{
We study a holographic gauge theory dual to the D3/D5 intersection. We consider a pure gauge B-field flux through the internal two-sphere wrapped by the probe D5--brane, which corresponds to a non-commutative configuration of adjoint scalars. There is a domain wall separating the theory into regions with different ranks of the adjoint group. At zero temperature the theory is supersymmetric and at finite temperature there is a critical point of a second order phase transition. We study the corresponding critical exponents and find that the second derivatives of the free energy, with respect to the bare mass and the magnetic field, diverge with a critical exponent of $-2/3$.

}


\subheader{}

\begin{document}
\maketitle

\section{Introduction}

Phase transitions are ubiquitous in nature and key to the understanding of a vast amount of phenomena, ranging from the description of boiling water to the comprehension of exotic states of matter such as quark-gluon plasma or quantum spin liquids. The modern classification distinguishes two major types of phase transitions:

First order phase transitions are characterised by the release or absorption of a latent heat. Typically these transitions are abrupt and powerful (large latent heat per volume). Across such a phase transition a first derivative of the thermodynamic potential has a discrete jump proportional to the latent heat. An important feature of the first order phase transitions is that their characteristics depend on the microscopic details of the system. 

The second order phase transitions (also called continuous phase transitions) are characterised by a diverging susceptibility (a second derivative of the free energy), scale invariance and infinite correlation length. The scale invariance washes out the microscopic details of the system allowing a classification of the continuous phase transitions into universality classes characterised entirely by the critical exponents of the divergent thermodynamic quantities. These remarkable properties make second order phase transition the subject of a considerable interest. 

In this paper we will uncover the existence of a second order critical point in a flavoured defect field theory holographically dual to the D3/D5--brane intersection. It is well known that at finite temperature this system exhibits a first order meson melting phase transition \cite{Myers:2006qr} (see also refs.~\cite{Babington:2003vm} and \cite{Albash:2006ew}). This meson melting transition is a close analogue of the second ordered confinement/deconfinement phase transition in Quantum Chromodynamics. It is believed that the first order of the meson melting transition in many holographic systems (including the D3/D5 system) is an artefact of the large $N$ limit. In the dual geometry this is represented by the existence of a topology changing merger transition of the probe D5--brane. In more details it is a transition between embeddings closing above the horizon of the geometry (referred to as Minkowski embeddings) and embeddings that reach all the way to the horizon (Black hole embeddings). 

Having mentioned the appealing properties of second order phase transitions it is natural to attempt to amend the holgraphic set-up so that the meson melting transition is continuous. One way to achieve this is to place the theory at a finite baryon density \cite{Kobayashi:2006sb}, corresponding to the introduction of a 10D electric field. Charge conservation requires a number of fundamental strings to be attached to the Mikowski probe D--branes to carry the charge into the AdS black hole, which is dynamically unstable. Remarkably, \cite{Kobayashi:2006sb} the probe D--brane develops throats/spikes mimicking fundamental strings and Mikowski embeddings are approximately realised within the black hole embeddings. The phase transition no longer corresponds to a topology change transition and a critical point of a second order phase transition naturally arises for a given parameter set.

A similar mechanism, which is a magnetic analogue of the baryon density case, was considered in ref.~\cite{Filev:2014mwa}, where a D5-brane probe in global AdS$_5\times S^5$ space-time was considered. The dual field theory has fundamental flavours confined to a maximal two-sphere within the three-sphere where the field theory lives. Introducing external magnetic field to the dual field theory corresponds to the introduction of a magnetic monopole \cite{Chunlen:2014zpa} in the world volume of the probe D5--brane, which prohibits embeddings wrapping a shrinking two-sphere within the AdS$_5$ subspace, unless D3--branes sourcing the monopole are attached. Remarkably these D3--branes are dynamically realised as throats by Mikowski D5-brane embeddings. Once again a topology change transition is avoided and a (quantum) critical point of a second order phase transition is realised. 

In this paper we consider yet another approach to avoid the topology change transition and this is to consider a pure gauge B--field flux through the internal $S^2$ wrapped by the D5--brane probe within the compact $S^5$ subspace. This set-up was first considered in refs.~\cite{Arean:2006vg,Arean:2007nh}, where the meson spectrum was obtained. Various aspects of this holographic set-up were studied in ref.~\cite{Myers:2008me} where it was argued that the magnetic flux corresponds to a non-commutative configuration of adjoint scalars. Furthermore, the fundamental fields are confined to a defect that serves as a domain wall separating the dual gauge theory into regions with different ranks of the gauge group. Extensive study of the phase diagram with a very general ansatz was considered in ref.~\cite{Wapler:2009rf}. Finally, quasinormal modes analysis at vanishing bare mass was performed in ref.~\cite{Itsios:2015kja}.

We focus on the holographic renormalization of the set-up, the existence of a second order phase transition and the properties of the theory near criticality. Interestingly we uncover the same critical exponents as in ref.~\cite{Filev:2014mwa}. We also revisit the supersymmetric regime of the theory generalizing it to the case of a non-vanishing bare mass and employing kappa symmetry to confirm the non-broken supersymmetry. The paper is organised as follows:
\begin{itemize}
\item In Section 2 we describe general properties of the holographic set-up with emphasis on Ramond-Ramond charge conservation and the existence of a domain wall separating regions with different ranks of the gauge group. 

\item In Section 3 we study the general supersymmetric embeddings. We use kappa symmetry to argue that all of the original supersymmetry of the D3/D5 system is preserved by the flux on the internal $S^2$. We construct supersymmetric D5--brane embeddings interpolating between a stack of half D3--brane embeddings parallel to the D3--branes sourcing the geometry and regular D5-brane embeddings. We discuss the field theory interpretation of the flux and how the domain wall is realised.

\item Section 4 studies the thermodynamics and holographic renormalization of the theory. We derive expressions for the fundamental condensates associated to the bare mass and the position of the defect. We discuss the phase structure of the theory with a focus on the existence of a critical point of a second order phase transition. We obtain numerically the critical exponents of the theory. 

\item Finally, we conclude with a brief discussion in Section 5.
\end{itemize}

\section{Holographic Set-Up}
\subsection{Probe configuration}\label{sec:probe_config}
In this section we outline the holographic set-up.\footnote{Note that this set-up has been first used in ref.~\cite{Myers:2008me}, where the transport properties of defect field theory have been explored.} We consider an AdS$_5\times S^5$ black hole supergravity background with metric:
\begin{eqnarray}\label{BH-metric}
ds^2 &=& -\frac{u^4-u_0^4}{u^2\,R^2}\,dt^2+\frac{u^2}{R^2}\left(dx_1^2+dx_2^2+dx_3^2\right) +\frac{u^2\,R^2}{u^4-u_0^4}du^2 + R^2\,d\Omega_5^2\ , \\
C_{(4)} &=& \frac{1}{g_s}\frac{u^4}{R^4}dx_0\wedge dx_1\wedge dx_2\wedge dx_3\ .\label{C4-bakcground}
\end{eqnarray}
Here $u_0$ is the radius of the black hole horizon related to the Hawking temperature of the background via $u_0 = \pi R^2\,T$ and $R$ is the radius of the internal $S^5$ given by $R^4 = 4\pi g_s N_c \alpha'^2$. 
It is convenient to use the following parametrisation of the unit 5-sphere:
\begin{eqnarray}
&&d\Omega_5^2 = d\theta^2+\cos^2\theta\,d\Omega_2^2 + \sin^2\theta\,d\tilde\Omega_2^2\ , \\
&&d\Omega_2^2=d\alpha^2+\sin^2\alpha \,d\beta^2\ ,~~~d\tilde\Omega_2^2=d\tilde\alpha^2+\sin^2\tilde\alpha \,d\tilde\beta^2\nonumber\ .
\end{eqnarray}
Next we consider a D5--brane embedding extended along the time direction $t$, two of the directions of the dual field theory $x_1$ and $x_2$ and localised along the third coordinate $x_3$\footnote{Note that the gauge field that we will turn on will force the D5--brane to develop a profile in the $u, x_3$ plane.}. The D5--brane also wraps the unit $S^2$ inside the $S^5$ part of the geometry parametrised by $\alpha$ and $\beta$. We let the D5--brane embedding develop a non-trivial profile $\theta(u)$ and $x_3(u)$. The dual defect field theory was first analysed in ref.~\cite{DeWolfe:2001pq} in the special case when the D5--branes have vanishing separation at infinity (corresponding to a zero bare mass) and the theory is superconformal. In this paper we are interested in a sort of deformation of that theory caused by turning on a non-trivial profile for the $U(1)$ gauge field on the $S^2\subset S^5$. We consider the following ansatz:
\begin{equation}\label{B-filed}
A_{(1)} = -\frac{H}{2\pi\alpha'}\,R^2\,\cos\,\alpha\,d \beta\ ,~~~F_{(2)} = dA_{(1)} = \frac{H}{2\pi\alpha'}\,R^2\,\sin\,\alpha\,d\alpha \wedge d \beta\ .
\end{equation}
One can show that the ansatz (\ref{B-filed}) is consistent with the equations of motion of the D5--brane. The full (DBI $+$ WZ) action of the D5--brane embedding is given by:
\begin{equation}\label{SD5_full}
S_{D5} = -\frac{N_f\mu_5}{g_s}\int\limits_{{\cal M}_6} d^6\xi\,e^{-\Phi}\sqrt{-|G_{ab}+{\cal F}_{ab}|}+N_f\mu_5\int\limits_{{\cal M}_6}{\cal P}\left[\sum_p C_{(p)}\wedge e^{\cal F}\right]\ ,
\end{equation}
where ${\cal F}_{(2)} =-{\cal P}[B_{(2)}] +2\pi\alpha' F_{(2)}$, $F_{(2)}$ is the field strength of the $U(1)$ gauge  field living on the D5--brane and $B_{(2)}$ is the Kalb-Ramond B-field. The main contribution\footnote{The $C_{(6)}$ Ramond-Ramond potential sourced by the D5--brane is suppressed by a a factor of $g_s$ which corresponds to a $1/N_c$ suppression. }  to the Wess-Zumino term in (\ref{SD5_full}) is given by:
\begin{equation}\label{WZ-action}
S_{WZ}=N_f\mu_5\int\limits_{{\cal M}_6} {\cal P}\left[C_{(4)}\wedge {\cal F}_{(2)}\right] =N_f\mu_5\int\limits_{{\cal M}_6} {\cal P}\left[C_{(4)}\right]\wedge {\cal F}_{(2)}\ .
\end{equation}
We will show that in addition to the leading order contribution from equation (\ref{C4-bakcground}) the background $C_{(4)}$ has an additional contribution from the D3--brane charge density supporting the magnetic monopole implied by the ansatz for the $U(1)$ gauge field~(\ref{B-filed}). 

\subsection{Charge Conservation.}
Let us consider the gauge transformation:
\begin{equation}\label{gauge_transformation}
C_{(4)} \to C_{(4)}+d\Lambda_{(3)}\ ,
\end{equation}
where $d\Lambda_{(3)}$ has legs along the D5--brane world volume.
For the variation of the WZ action (\ref{WZ-action}) we obtain:
\begin{equation}\label{3d-bry}
\delta S_{WZ}=N_f\mu_5\int\limits_{{\cal M}_6}	d\Lambda_{(3)}\wedge {\cal F}_{(2)} =N_f\mu_5\int\limits_{{\cal M}_6}	\Lambda_{(3)}\wedge d{\cal F}_{(2)}= 4\pi H R^2\,N_f\mu_5\int\limits_{{\cal M}_3}	\Lambda_{(3)}\ ,
\end{equation}
where ${\cal M}_3$ is a three dimensional slice of the D5--brane world volume at the point $u_{min}$ where the internal $S^2$ (parametrised by $\alpha$ and $\beta$) shrinks to zero radius\footnote{In obtaining (\ref{3d-bry}) we have used that $d{\cal F}_{(2)}=2\,H R^2\,\delta(u-u_{min})\,\sin\alpha\,du\wedge d\alpha\wedge d\beta$ with the convention $\int\limits_{u_{min}}^{u_{min}+\epsilon}du\, \delta(u-u_{\min}) = 1/2$.}.

As one can see from equation (\ref{3d-bry}) there is a three dimensional boundary in the world volume of the D5--brane. In fact, this is the magnetic monopole (in six dimensions) supporting the ansatz (\ref{B-filed}). It is a standard result in D--brane dynamics that magnetic monopole in the world volume of a Dp--brane is sourced by a D(p-2)--brane ending on the Dp--brane. In our case we need to attach certain number $n$ of D3--branes to the boundary ${\cal M}_3$ (with the appropriate orientation) to cancel the variation of the WZ action in (\ref{3d-bry}). Indeed, under the gauge transformation (\ref{gauge_transformation}) the WZ action of this additional D3--branes would transform as:
\begin{equation}
\delta S^{D3}_{WZ}= -k\,\mu_3\int\limits_{{\cal M}_3}	\Lambda_{(3)}\ .
\end{equation}  
It is straightforward to verify that the condition $\delta S_{WZ} + \delta S^{D3}_{WZ} = 0$ is equivalent to:
\begin{equation}\label{flux_quant}
\Phi_H = 4\pi N_f\left(\frac{H R^2}{2\pi\alpha'}\right) =2\pi\, k\ ,
\end{equation}
that is the flux of the $U(1)$ field through the unit $S^2$ is quantised according to the Aharonov-Bohm effect. 

Even though we have ensured charge conservation at the boundary ${\cal M}_3$, we now have the problem of attaching the other end of the stack of n D3--branes. We have several options: we could let the new stack of D3--branes carry the charge to infinity, we could attach the new stack to the original stack of D3--branes (most likely unstable) or if at finite temperature we could let the new stack of D3--branes carry the charge to the horizon of the geometry. Note that these are only kinematic considerations and our goal is not to engineer any of the above brane configurations. Instead, in the next sections, we analyse the dynamics of the D5--brane probe and show that the additional stack of D3--branes is dynamically realised as a throat in the D5-brane worldvolume. 

At zero temperature there are supersymmetric solutions, generalizing the ones obtained in ref.~\cite{Myers:2008me} and interpolating between D5--brane embeddings at large radial distance and ``natural'' (parallel to the stack sourcing the geometry) D3--brane probes. Note that in addition to solving the charge conservation problem, this introduces a domain wall separating regions with different ranks of the gauge group~\cite{Myers:2008me}. 

At finite temperature the only complete solutions that we uncover are black hole D5--brane embeddings. The Minkowski phase is realised within the black hole one with the D5--brane embeddings developing D3--brane throats carrying the charge to the horizon.

\section{The SUSY case. A Domain Wall.}
In this subsection we focus on the zero temperature case ($u_0=0$ in equation (\ref{BH-metric})). The massless case was first analysed in ref.~\cite{Myers:2008me}. In the following we will consider the general case (allowing a non-zero bare mass of the defect fields) and will supplement these studies with a kappa symmetry derivation of the supersymmetric embeddings. We begin by considering the effective action of the probe brane.
\subsection{Effective Action}
It is convenient to rewrite the metric of AdS$_5\times S^5$ as:
\begin{equation}\label{l-rho-metric}
ds^2 =\frac{\rho^2+l^2}{R^2}\left(-dt^2 + dx_1^2 + dx_2^2 + dx_3^2\right)+\frac{R^2}{\rho^2+l^2}\left(d\rho^2+\rho^2 d\Omega_2^2+dl^2+l^2d\tilde\Omega_2^2\right)\ ,
\end{equation}
where the connection to the coordinates in equation (\ref{BH-metric}) is given by: $l = u\,\sin\theta$ and $\rho = u\,\cos\theta$. Now considering an ansatz: $l(\rho)$, $x_3(\rho)$ and $A_\mu = 0$ we obtain: 
\begin{eqnarray}
{\cal L} &=& -\sin\alpha\,\sqrt{\rho^4+H^2(\rho^2+l(\rho)^2)^2}\sqrt{1+l'(\rho)^2 +\left(\frac{\rho^2+l(\rho)^2}{R^2}\right)^2 x_3'(\rho)^2}  + \nonumber \\
&&+\sin\alpha\,H\,R^2\left(\frac{\rho^2+l(\rho)^2}{R^2}\right)^2 x_3'(\rho)\ .
\end{eqnarray}
Note that $x_3(\rho)$ is a cyclic variable and can be eliminated by a Legender transformation:
\begin{equation}\label{LegSUSY}
\tilde{\cal L} ={\cal L} -\frac{\partial{\cal L}}{\partial{x_3'(\rho)}}x_3'(\rho) =-\sin\alpha\,\rho^2\,\sqrt{1+l'(\rho)^2}\ ,
\end{equation}
where we have used that $\partial{\cal L}/\partial{x_3'(\rho)}={\rm const} = 0$\footnote{Note that the choice $\partial{\cal L}/\partial{x_3'(\rho)}=0$ for the constant of integration is dictated by supersymmetry.} and
\begin{equation}
x_3'(\rho)=\pm\frac{H\,R^2}{\rho^2}\ .
\end{equation}
Remarkably, the Legender transformed Lagrangian (\ref{LegSUSY}) is exactly the same as in the supersymmetric $H=0$ case and the regular solution to the equation of motion for $l$ is given by $l(\rho) = m = {\rm const}$. In fact, in the next section we will use kappa symmetry to show that the solution:
\begin{equation}\label{SUSY_SOL}
x_3'(\rho)=\frac{H\,R^2}{\rho^2}\ , ~~~~l(\rho) = m = {\rm const}\ .
\end{equation}
is supersymmetric. 

\subsection{Kappa Symmetry}
The kappa symmetry matrix is given by:
\begin{equation}\label{Gamma_kappa}
\Gamma_\kappa =\frac{\sqrt{-|G_{ab}|}}{\sqrt{-|G_{ab}+{\cal F}_{ab}|}}\sum_{n=0}^\infty\frac{(-1)^n}{2^n n!}\gamma^{a_1 b_1\dots a_n b_n}{\cal F}_{a_1 b_1}\dots{\cal F}_{a_n b_n}\,\Gamma_{(0)}\otimes \sigma_3^{\frac{p-3}{2}-n}(i\sigma_2)\ ,
\end{equation}
where $\Gamma_{(0)}$ is given by:
\begin{equation}
\Gamma_{(0)} =\frac{1}{\sqrt{-|G_{ab}|}}\frac{1}{(p+1)!}\varepsilon^{a_1\dots a_{p+1}}\gamma_{a_1}\dots\gamma_{a_{p+1}}\ .
\end{equation}
Note that the matrices $\gamma_a$ are the pull back for the ``flat'' gamma matrices given by:
\begin{equation}
\gamma_a = \partial_a X^\mu\,E_\mu^{\bar \mu}\,\Gamma_{\bar\mu} ,
\end{equation}
where $E_\mu^{\bar \mu}$ are the tetrads of the metric (\ref{l-rho-metric}) and $\Gamma_{\bar\mu}$ are the flat gamma matrices associated to the tetrads. Note also that the matrices $\gamma^a$ are obtained by lifting the index of $\gamma_a$ with the inverse of the induced metric.

Next we proceed by considering the ansatz $l(\rho)$ and $x_3(\rho)$, the goal is to obtain the solution (\ref{SUSY_SOL}) from the requirement to preserve kappa symmetry. With this ansatz we obtain:
\begin{equation}\label{Gamma0}
\Gamma_{(0)} = \Gamma_{\bar\alpha\bar\beta}\,\frac{\left(\Gamma_{\bar\rho\bar x_3} + l'(\rho)\Gamma_{\bar l\bar x_3}+\frac{\rho^2+l(\rho)^2}{R^2} x_3'(\rho)\right)}{\sqrt{1+l'(\rho)^2 +\left(\frac{\rho^2+l(\rho)^2}{R^2}\right)^2 x_3'(\rho)^2}}\,\Gamma_{\bar t\,\bar x_1\bar x_2 \bar x_3}\ .
\end{equation}
Now using the ansatz for the gauge field (\ref{B-filed}) we obtain:
\begin{equation}
\Gamma_\kappa = \frac{\Gamma_{(0)}\otimes \sigma_3(i\sigma_2)-H\,\frac{\rho^2+l(\rho)^2}{\rho^2}\,\Gamma_{\bar\alpha\bar\beta}\,\Gamma_{(0)}\otimes (i\sigma_2)}{\sqrt{1+H^2\left(\frac{\rho^2+l(\rho)^2}{\rho^2}\right)^2}}
\end{equation}
Substituting $\Gamma_{(0)}$ from equation (\ref{Gamma0}) we obtain:
\begin{equation}
\Gamma_\kappa = \frac{\Gamma_{\bar\rho\bar x_3} +l'(\rho)\Gamma_{\bar l\bar x_3}+\frac{\rho^2+l(\rho)^2}{R^2}x_3'(\rho)}{\sqrt{1+l'(\rho)^2+\left(\frac{\rho^2+l(\rho)^2}{R^2}\right)^2 x_3'(\rho)^2}}.\frac{H\frac{\rho^2+l(\rho)^2}{\rho^2} +\Gamma_{\bar\alpha\bar\beta}\otimes \sigma_3}{\sqrt{1+H^2\left(\frac{\rho^2+l(\rho)^2}{\rho^2}\right)^2}}\Gamma_{\bar t\,\bar x_1\bar x_2 \bar x_3}\otimes i\sigma_2
\end{equation}
Now using that the Killing spinor $\varepsilon$ satisfies:\footnote{Note that equation (\ref{Killing}) is equivalent to the kappa symmetry condition for a natural D3--brane probe, since $\Gamma_{\bar t\,\bar x_1\bar x_2 \bar x_3}\otimes i\sigma_2$ is precisely the kappa matrix for such a probe which can be seen by using (\ref{Gamma_kappa}) with $p=3$.}
\begin{equation}\label{Killing}
\Gamma_{\bar t\,\bar x_1\bar x_2 \bar x_3}\otimes i\sigma_2 \varepsilon = \varepsilon\ ,
\end{equation}
we obtain the kappa symmetry condition:
\begin{equation}\label{kappa_1}
\Gamma_\kappa\,\varepsilon = \frac{\Gamma_{\bar\rho\bar x_3} +l'(\rho)\Gamma_{\bar l\bar x_3}+\frac{\rho^2+l(\rho)^2}{R^2}x_3'(\rho)}{\sqrt{1+l'(\rho)^2+\left(\frac{\rho^2+l(\rho)^2}{R^2}\right)^2 x_3'(\rho)^2}}.\frac{H\frac{\rho^2+l(\rho)^2}{\rho^2} +\Gamma_{\bar\alpha\bar\beta}\otimes \sigma_3}{\sqrt{1+H^2\left(\frac{\rho^2+l(\rho)^2}{\rho^2}\right)^2}}\,\varepsilon =\varepsilon\ .
\end{equation}
Note that if we set $l'(\rho) = 0$, $x_3'(\rho) = 0$ and $H=0$, equation (\ref{kappa_1}) reduces to:
\begin{equation}\label{projection2}
\Gamma_{\bar\rho\bar x_3\bar\alpha\bar\beta}\otimes \sigma_3 \,\varepsilon = \varepsilon\ ,
\end{equation}
which reduces the supersymmetry of the background by half. Now let us restore a non trivial $x_3'(\rho)$ and a non vanishing $H$ while keeping the projection (\ref{projection2}). We obtain:
\begin{equation}
\Gamma_\kappa\,\varepsilon = \frac{\Gamma_{\bar\rho\bar x_3} +\frac{\rho^2+l(\rho)^2}{R^2}x_3'(\rho)}{\sqrt{1+\left(\frac{\rho^2+l(\rho)^2}{R^2}\right)^2 x_3'(\rho)^2}}.\frac{H\frac{\rho^2+l(\rho)^2}{\rho^2} -\Gamma_{\bar\rho\bar x_3}}{\sqrt{1+H^2\left(\frac{\rho^2+l(\rho)^2}{\rho^2}\right)^2}}\,\varepsilon\ ,
\end{equation}
where we used that $\Gamma_{\bar\alpha\bar\beta}\otimes \sigma_3 \,\varepsilon = -\Gamma_{\bar\rho\bar x_3}\,\varepsilon$. It is easy to check that only for $x_3'(\rho)$ given in equation (\ref{SUSY_SOL}) we have:
\begin{equation}
\Gamma_\kappa\,\varepsilon = \frac{(H\frac{\rho^2+l(\rho)^2}{\rho^2} +\Gamma_{\bar\rho\bar x_3})(H\frac{\rho^2+l(\rho)^2}{\rho^2} -\Gamma_{\bar\rho\bar x_3})}{1+H^2\left(\frac{\rho^2+l(\rho)^2}{\rho^2}\right)^2}\,\varepsilon\ = \varepsilon
\end{equation}
and kappa symmetry is preserved without additional projection. Therefore, the solution (\ref{SUSY_SOL}) preserves ${\cal N}=2$ supersymmetry and the corresponding dual field theory is given by a generalization of the Lagrangian proposed in ref.~\cite{DeWolfe:2001pq} to a non-zero bare mass. 

\subsection{Domain Wall}
In this section we show that the solutions constructed above describe a domain wall separating the Minkowski space into regions with different gauge group. Furthermore, for massless embeddings the theory remains conformal. Let us integrate the equation for $x_3$ in (\ref{SUSY_SOL}). We obtain:
\begin{equation}\label{SUSYx3}
x_3(\rho) = x_3(\infty) -\frac{H\,R^2}{\rho}\ .
\end{equation}
One can see that for large $\rho$ $x_3$ approaches the  constant value $x_3(\infty)$ and the solution asymptotes to the ``usual'' D5-brane embedding (the one with $H=0$) on the other hand in the limit $\rho\to 0$ we have $x_3\to -\infty$, while the internal $S^2$ shrinks to zero radius. Therefore, at $\rho \to 0$ the solution looks like a D3--brane spanned along the $t, x_1, x_2$ and $x_3$ directions. We can estimate the number of these D3--branes. Let us evaluate the contribution to the DBI and WZ actions of the probe D5--brane of the half Minkowsi space ${\cal M}_4^{(-)}$ spanning the $t,x_1, x_2$ directions and the ray $-\infty < x_3 < -\Lambda$ with $\Lambda\gg 1$. To leading order we obtain:
\begin{eqnarray}
\Delta S_{D5}^{DBI} &=& -N_f\mu_5\,4\pi\,H \frac{m^4}{R^2}\int\limits_{{\cal M}_4^{(-)}} d^4 x \label{effective-D3-DBI}\\
\Delta S_{D5}^{WZ} &=& N_f\mu_5\,4\pi\,H \frac{m^4}{R^2}\int\limits_{{\cal M}_4^{(-)}} d^4 x\ .\label{effective-D3-WZ}
\end{eqnarray}
Comparing equations (\ref{effective-D3-DBI}) and (\ref{effective-D3-WZ}) to the contribution to the DBI and WZ actions of a probe D3--brane (positioned at $\rho =0, l = m$) of the same half-space:
\begin{eqnarray}
\Delta S_{D3}^{DBI} &=& -\mu_3\,k \frac{m^4}{R^4}\int\limits_{{\cal M}_4^{(-)}} d^4 x \\
\Delta S_{D3}^{WZ} &=& \mu_3\,k \frac{m^4}{R^4}\int\limits_{{\cal M}_4^{(-)}} d^4 x\ ,
\end{eqnarray}
we arrive at equation (\ref{flux_quant}) relating the number of effective D3--branes to the flux of the $U(1)$ field. Therefore, the D5--brane embedding interpolates between the solutions analysed in ref.~\cite{DeWolfe:2001pq} and a stack of $k$ D3--branes parallel to the stack of D3--branes sourcing the geometry. 

\begin{figure}[t]
   \centering
   \includegraphics[width=7cm]{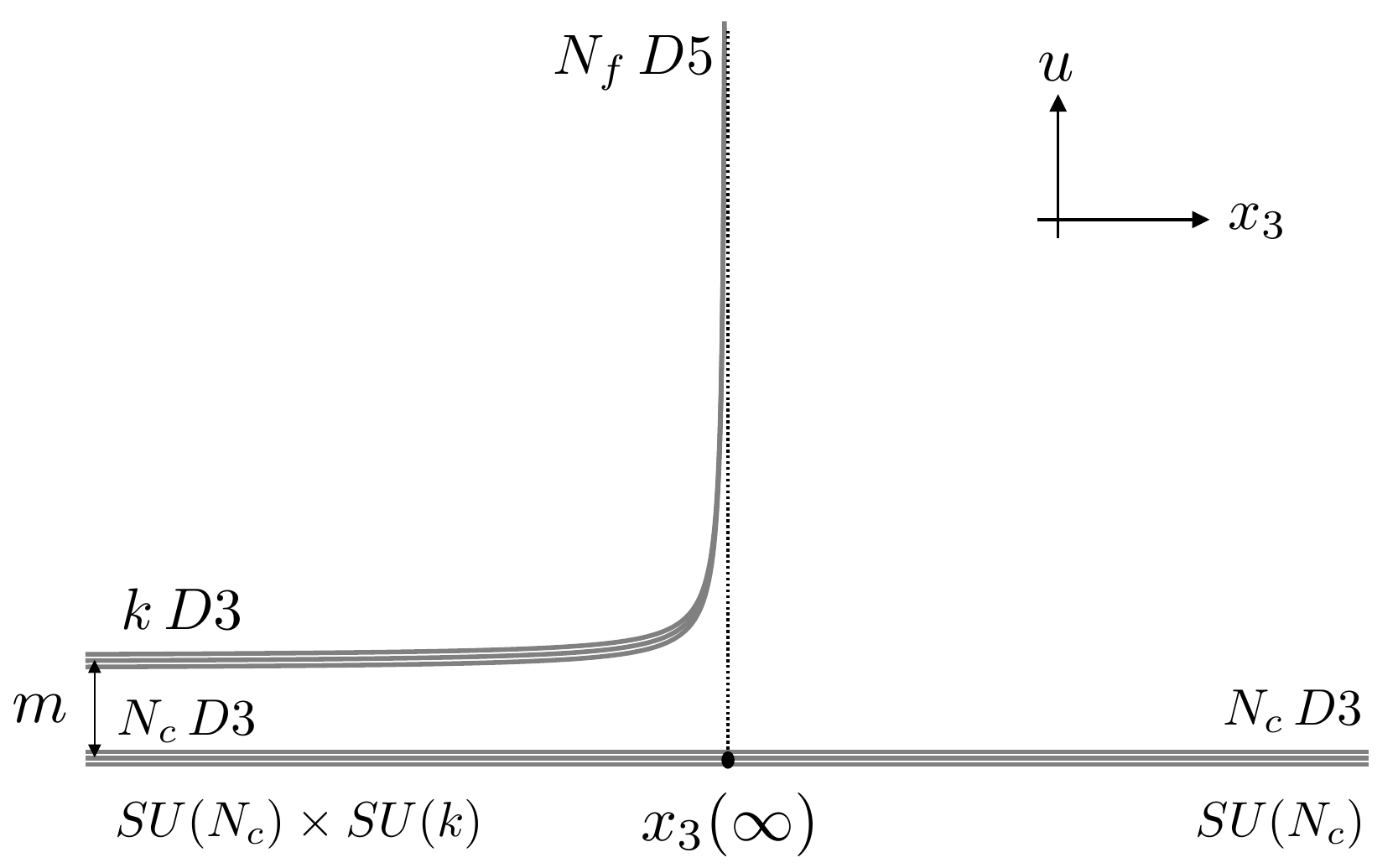}
   \includegraphics[width=7cm]{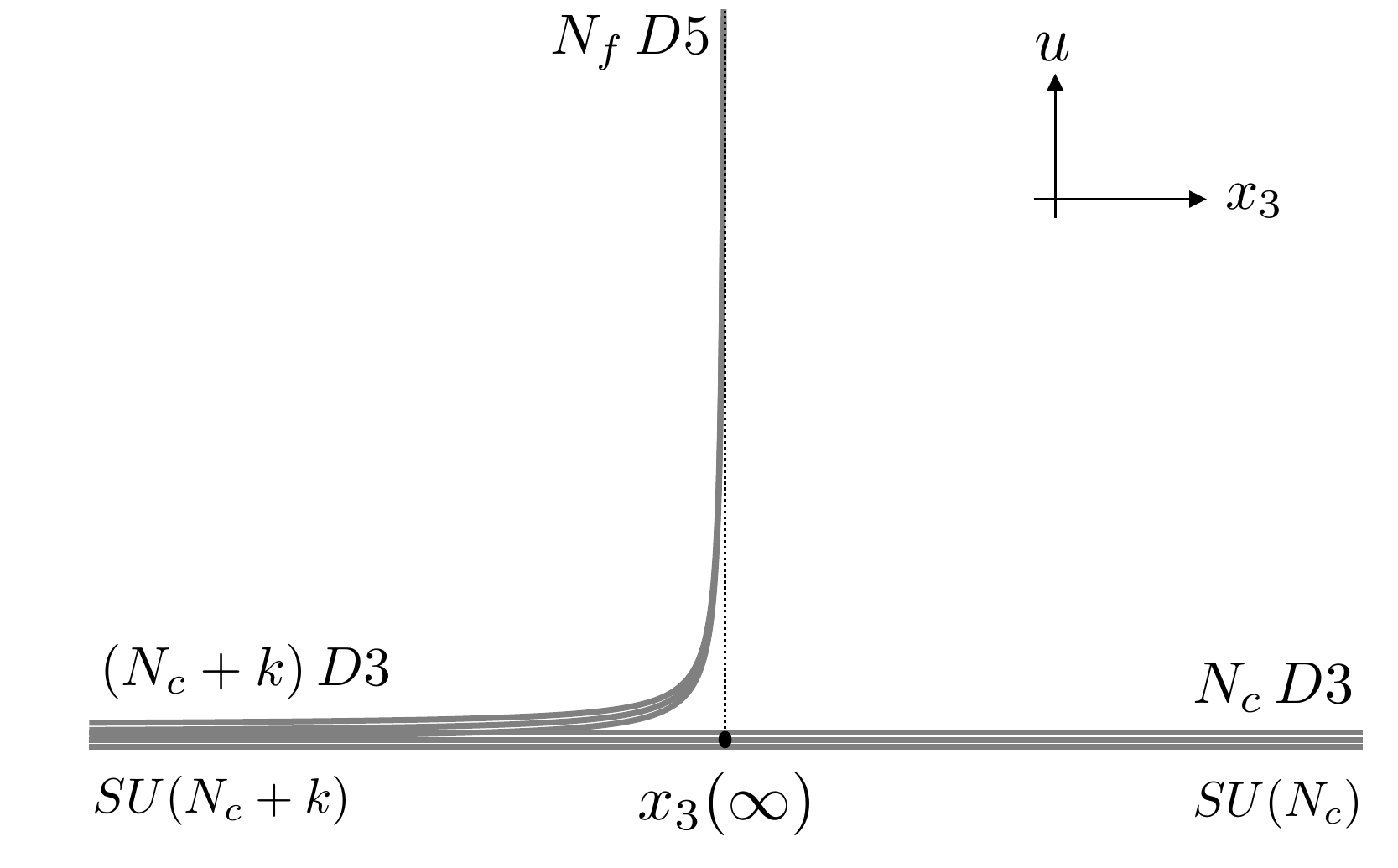}
   \caption{{\it Left side:} A plot of massive D5--brane embeddings. The embeddings start as regular D5--brane embeddings at large $u$ and approximate a stack of k D3--brane embeddings for $x_3 \to -\infty$. {\it Right side:}  A plot of massless D5--brane embeddings. The stack of k D3--branes overlaps with the original stack of D3--branes for $x_3 \to -\infty$.}
   \label{fig:dw_susy}
\end{figure}

In figure~\ref{fig:dw_susy} we have presented plots of the solutions in the $x_3, u$ plane. The left plot represents a solution with finite mass $m$. The solution start as a regular D5--brane embedding at large $u$ and bends towards the negative $x_3$ axes as $u\to m$. As equations (\ref{effective-D3-DBI}) and (\ref{effective-D3-WZ}) suggest for $u\to m$ and $x_3\to-\infty$ the D5--brane embeddings approximate a stack of $k$ D3--branes. As a results in the dual field theory there is a domain wall (sitting at $x_3(\infty)$) splitting the Minkowski space into regions with different gauge groups. For $x_3 > x_3(\infty)$ the gauge group is the original $SU(N_c)$, while for $x_3 < x_3(\infty)$ the gauge group is enhanced to $SO(N_c)\times SO(k)$ due to the additional D3--brane charge. The right plot in figure \ref{fig:dw_susy} represents massless ($m=0$) D5--brane embedding. In this case the effective stack of $k$ D3--branes overlaps with the original D3--branes. This enhances the gauge group on the left side ($x_3 < x_3(\infty)$) of the domain wall to the full $SU(N_c+k)$ group. Furthermore, one can check \cite{Myers:2008me} that when $m=0$ the induced metric of the D5--brane can be written as:
\begin{equation}
d\gamma^2=\frac{\tilde\rho^2}{(1+H^2)\,R^2}(-dt^2+dx_1^2+dx_2^2)+\frac{(1+H^2)\,R^2}{\tilde\rho^2}d\tilde\rho^2 +R^2d\Omega_2^2\ ,
\end{equation}
where $\tilde\rho = \sqrt{1+H^2}\rho$. This is the metric of an AdS$_4\times S^2$ space time (with different radii of the AdS$_4$ and $S^2$ manifolds). This suggests that for $m=0$ the classical conformal symmetry of the dual theory is not broken by the domain wall. 

We can use the ${\cal N}=2$ supersymmetry to generalize the field theory interpretation of ref.~\cite{Myers:2008me} to the non-zero mass case. Namely, that the flux (\ref{flux_quant}) carried by the D5--branes corresponds to producing a noncommutative configuration of adjoint scalars in a $SU(k)$ subgroup of the $SU(N_c) \times SU(k)$ gauge group on the relevant side of the domain wall. Now using supersymmetry one can match the profile of the D5-branes to the profile of the adjoint scalars $X_A$, one has :
\begin{equation}\label{profileCFT}
\frac{1}{N_f}\sum\limits_{A=1}^6 {\rm Tr}(X_A^2)=\frac{m^2+\rho^2}{(2\pi\alpha')^2} = m_q^2 + \frac{k^2}{4N_f^2}\frac{1}{x_3^2}\ ,
\end{equation}
where we have used equations (\ref{flux_quant}) and (\ref{SUSYx3}) and for simplicity we have set the position of the domain wall $x_3(\infty)$ to zero. We have also used that $m_q = m/(2\pi\alpha')$. Clearly for $m_q =0$ equation (\ref{profileCFT}) reduces to the one given in ref.~\cite{Myers:2008me} and the gauge group on one side of the domain wall is promoted to $SU(N_c + k)$.

\section{Thermodynamics and Holographic Renormalization} 
Various properties of the termodynamics of the D3/D5 system have been studied in refs.~\cite{Mateos:2007vn}, \cite{Wapler:2009rf} and \cite{Filev:2009xp, Grignani:2012jh, Grignani:2014vaa, Georgiou:2015pia, Grignani:2016npu, Bea:2016fcj, Alho:2016gdf,Conde:2016hbg, Penin:2017lqt,Grignani:2018qrn}. In ref.~\cite{Wapler:2009rf} a very general anstaz has been considered and the author has uncovered a number of interesting properties of the system. In this work we uncover the existence of a critical point of a second order phase transition drawing a parallel with the studies of ref.~\cite{Filev:2014mwa}.

\subsection{Holographic Renormalization}
We begin by writing the Lagrangian corresponding to the effective action (\ref{SD5_full}) for the background and ansatz given in equations (\ref{BH-metric})-(\ref{B-filed}). In fact, we will work with the Euclidean version of the Lagrangian, which in our case differ by an overall minus sign. Furthermore, we define:
\begin{equation}
\tilde u = u/u_0;~~~\tilde R^2 = R^2/u_0;
\end{equation}
and consider the dimensionless Lagrangian:
\begin{equation}
{\cal L}_E = \tilde u^2\sqrt{H^2+\cos^4\theta(\tilde u)}\left(1+\frac{\tilde u^4-1}{\tilde R^4}\,x_3'(\tilde u)^2+\frac{\tilde u^4-1}{\tilde u^2}\,\theta'(\tilde u)^2\right)^{1/2}-\frac{H\,\tilde u^4}{\tilde R^2}\,x_3'(\tilde u)\ .
\end{equation}
Note that $x_3$ is a cyclic variable and hence $\frac{\partial {\cal L}_E}{\partial x_3'(u)} = {\rm const}$. One can show that the following choice for the constant of integration:
\begin{equation}
\frac{\partial {\cal L}_E}{\partial x_3'(\tilde u)} = -\frac{H}{\tilde R^2}
\end{equation}
is the only one leading to finite Lagrangian density and embeddings consistent with charge conservation. With this choice it is easy to obtain:
\begin{equation}\label{x3-T}
x_3'(\tilde u) = \frac{H \tilde R^2\left(1+\frac{\tilde u^4-1}{\tilde u^2}\,\theta'(\tilde u)^2\right)^{1/2}}{\sqrt{H^2+\tilde u^4\cos^4\theta(\tilde u)}}\ ,
\end{equation}
where to be consistent with the zero temperature limit we have chosen the positive sign for $x_3'(u)$. Using (\ref{x3-T}) it is straightforward to to write down the Legender transform of ${\cal L}_E$:
\begin{equation}\label{Leg-Lag}
\tilde{\cal L}_E = {\cal L}_E - \frac{\partial {\cal L}_E}{\partial x_3'(\tilde u)}x_3'(\tilde u)= \sqrt{H^2+\tilde u^4\cos^4\theta(\tilde u)}\left(1+\frac{\tilde u^4-1}{\tilde u^2}\,\theta'(\tilde u)^2\right)^{1/2}\ .
\end{equation}
Solving perturbatively at large $u$ the equation of motion for $\theta(u)$ derived from (\ref{Leg-Lag}), one obtains:
\begin{equation}\label{asymp-sin th}
\theta(\tilde u) = \frac{\tilde m}{\tilde u}+\frac{\tilde c}{\tilde u^2}+\dots\ .
\end{equation}
Using (\ref{asymp-sin th}) it is easy to show that there are no new divergences introduced by the $U(1)$ gauge field. Therefore, we can regularize the Euclidean actions corresponding to ${\cal L}_E$ and $\tilde {\cal L}_E$ using the approach of \cite{Myers:2006qr}. For D5--brane embeddings reaching the horizon (Black Hole embedding) we can write:
\begin{eqnarray}
I_E &=& \int\limits_{1}^\infty d\tilde u\,\left({\cal L}_E +\frac{dI_{sub}(\theta(\tilde u),\tilde u)}{d\tilde u}\right) +I_{sub}(\theta(1),1)\ ,\\
\tilde I_E &=& \int\limits_{1}^\infty d\tilde u\,\left(\tilde {\cal L}_E +\frac{dI_{sub}(\theta(\tilde u),\tilde u)}{d\tilde u}\right) +I_{sub}(\theta(1),1)\ \label{leg-action},\\
I_{sub}(\theta, \tilde u) &=& -\frac{\tilde u^3}{3}-\frac{\tilde u^3}{2}\sin^2\theta\ .
\end{eqnarray}
The corresponding thermodynamic potentials (or rather their two dimensional densities) are given by:
\begin{eqnarray}
F &=& \frac{4\pi \mu_5 N_f}{g_s}\,u_0^3 I_E =\sqrt{\frac{\lambda}{2}}N_f N_c \,T^3{I_E}\ , \\
\tilde F &=& \sqrt{\frac{\lambda}{2}} N_f N_c\, T^3{\tilde I_E}\ .\label{legF}
\end{eqnarray}
We identify $F$ as the free energy density satisfying:
\begin{equation}
dF = - S dT +C_m dm_q + C_{\Delta x_3} d \Delta x_3\ ,
\end{equation}
where: $S$ is the entropy density, $T$ is the temperature, $C_m$ is the fundamental condensate, $m_q = u_0\,\tilde m/(2\pi\alpha')$ is the fundamental mass, $C_{\Delta x_3}$ is the condensate of the operator obtained by varying the defect field theory action with respect to its position along $x_3$ and $\Delta x_3 \equiv x_3(\infty)-x_3(u_0)$. Note that since $x_3$ is defined up to an additive constant we can always define $x_3(u)$ to satisfy $x_3(u_0)=0$ and have $\Delta x_3 = x_3(\infty)$.

To calculate $C_{\Delta x_3}$ we vary the free energy density $F$ at fixed $T$ and $m$, which is equivalent of keeping fixed $u_0$ and $\theta(u)$. It is easy to obtain:
\begin{eqnarray}
\delta F = \sqrt{\frac{\lambda}{2}}N_f N_c T^3\int\limits_{1}^\infty d\tilde u\, \frac{\partial}{\partial \tilde u}\left(\frac{\partial{\cal L}_E}{\partial x_3'(\tilde u)}\delta x_3(\tilde u)\right)=-\sqrt{\frac{\lambda}{2}} N_f N_c \pi H\, T^4\delta \Delta x_3\ ,
\end{eqnarray}
Therefore,
\begin{equation}\label{pos_cond}
C_{\Delta x_3} \equiv \left(\frac{\delta F}{\delta \Delta x_3}\right)_{m,T}=-\sqrt{\frac{\lambda}{2}}N_f N_c \pi H\, T^4\ .
\end{equation}
To relate the fundamental condensate to the supergravity parameters it is convenient to consider the ``canonical'' free energy $\tilde F$ given in (\ref{legF}). It is easy to show that the following relations hold:
\begin{eqnarray}
\tilde F &=& F - \Delta x_3\,C_{\Delta x_3}\ ,\ \\
d \tilde F &=&-SdT + C_m dm_q - \Delta x_3\,dC_{\Delta x_3}\ .\label{tildeF}
\end{eqnarray}
Now fixing $T$ and $C_{\Delta x_3}$ and varying $m$, that is fixing $u_0$ and $H$ and varying $\theta(u)$ with large $u$ expansion $\delta \theta(u) = \delta m/u + O(1/u^2)$, we get:
\begin{eqnarray}
\delta \tilde F &=& \sqrt{\frac{\lambda}{2}}N_f N_c T^3\left[\int\limits_{1}^\infty d\tilde u\frac{\partial}{\partial \tilde u}\left(\frac{\partial\tilde {\cal L}_E}{\partial\theta'(\tilde u)}\delta \theta(\tilde u)\right) +\frac{\partial I_{sub}}{\partial \theta(1)}\delta\theta(1)\right] =  \\
&=&-\sqrt{\frac{\lambda}{2}}N_f N_c T^3\,\tilde c\,\delta \tilde m\ =-N_f N_c\,T^2\tilde c\, \delta m_q,\nonumber
\end{eqnarray}
where we used that $\tilde m = 2\pi\alpha' m_q/u_0$.
Therefore,
\begin{equation}\label{fund_cond}
C_m \equiv \left(\frac{\delta \tilde F}{\delta m_q}\right)_{C_{\Delta x_3},T} = -N_f N_c T^2\tilde c\ .
\end{equation}
\subsection{Phase structure}
\begin{figure}[t]
   \centering
   \includegraphics[width=12cm]{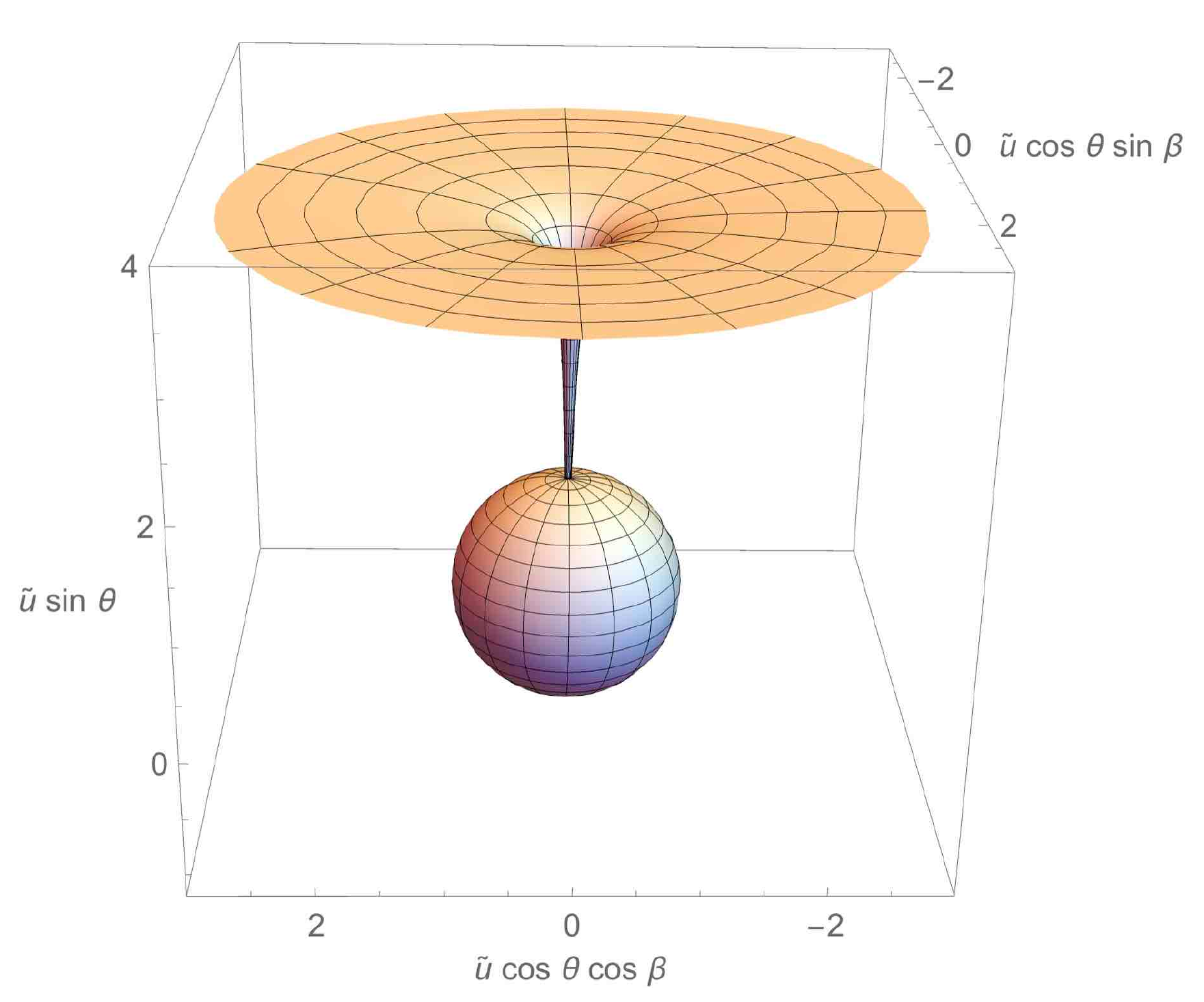}
   \caption{A plot of a Black hole embedding mimicking a Minkowski embedding connected with a D3-brane throat to the horizon. The dashed vertical line represent the critical parameter $\tilde m_{cr}$ at which the phase transition takes place.}
   \label{fig:Throat}
\end{figure}

In this section we solve numerically the EOM for $\theta(\tilde u)$ obtained by varying the Lagrangian (\ref{Leg-Lag}) and solve for $x_3(\tilde u)$ using equation (\ref{x3-T}). The standard classification of the possible D5--brane embeddings \cite{Myers:2006qr} is based on the their topology. In particular one distinguishes Black Hole embeddings characterised by reaching the horizon and Minkowski embeddings which close above the horizon by having the internal $S^2$ wrapped by the brane shrink to zero size. However, as we showed in section \ref{sec:probe_config} charge conservation demands that a number of D3--branes is attached to the D5--branes at the point where the $S^2$ shrinks to support the required magnetic monopole. This observation renders the class of Minkowski embeddings incomplete since no stable configurations with attached D3--branes carrying away the Ramond-Ramond charge were identified. Instead, the Minkowski embeddings develop a narrow throat connecting them to the horizon of the geometry and become a subset of the Black hole embeddings. In figure~\ref{fig:Throat} we have presented a 3D plot of such an embedding, where the horizon is represented by a two-sphere parametrised by $\theta$ and $\beta$, while the internal $S^2$ is represented by an $S^1$ parametrised by $\beta$. One can see that the Black hole embedding mimics a Minkowski embedding connected with a D3--brane throat to the horizon. 

Our next step is to explore the behaviour of the fundamental condensate $C_m$ given in equation (\ref{fund_cond}) as a function of the bare mass $m_q$ at a fixed condensate $C_{\Delta x_3}$ defined in equation (\ref{pos_cond}). 

\begin{figure}[h]
   \centering
   \includegraphics[width=6cm]{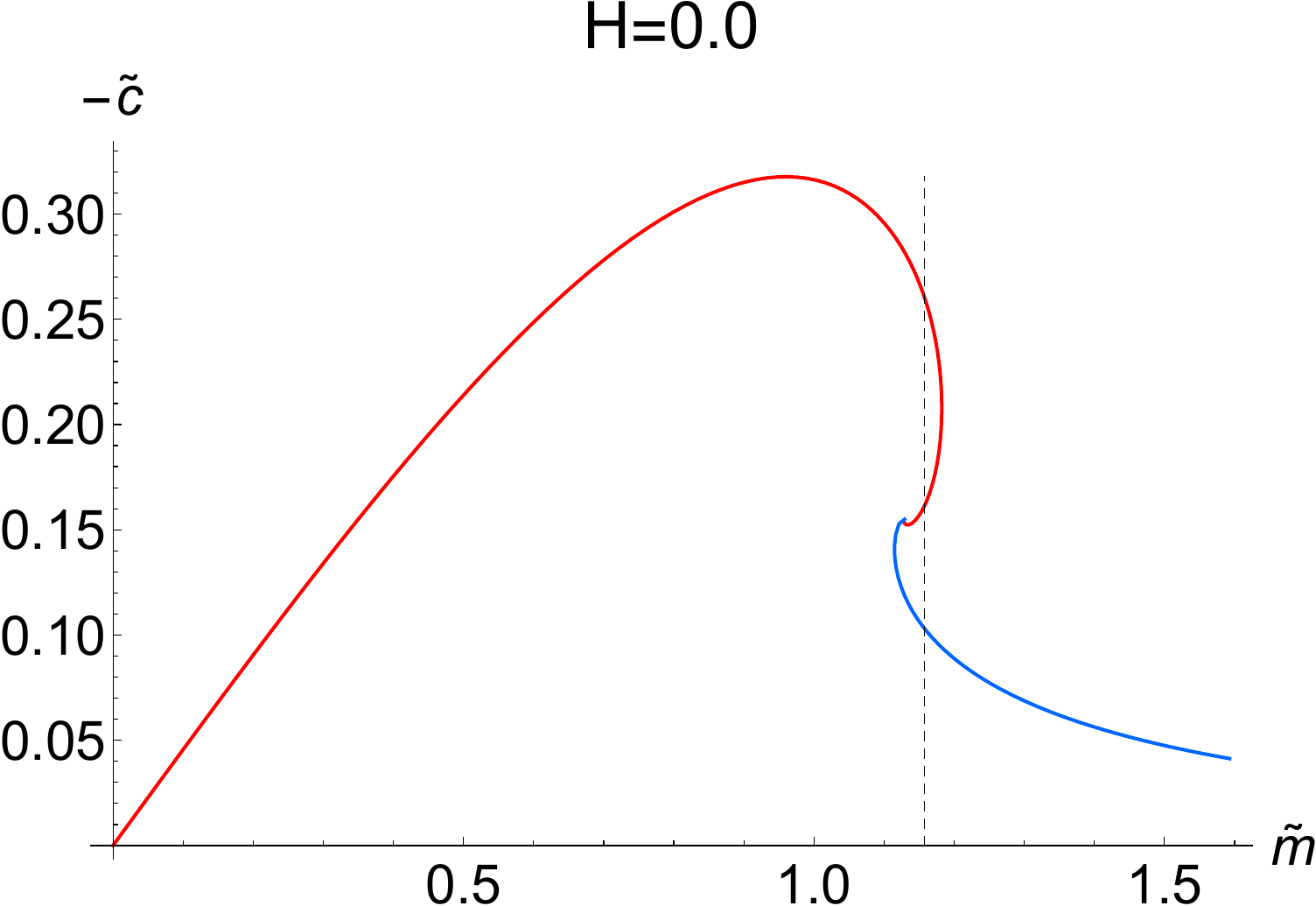}
   \includegraphics[width=6cm]{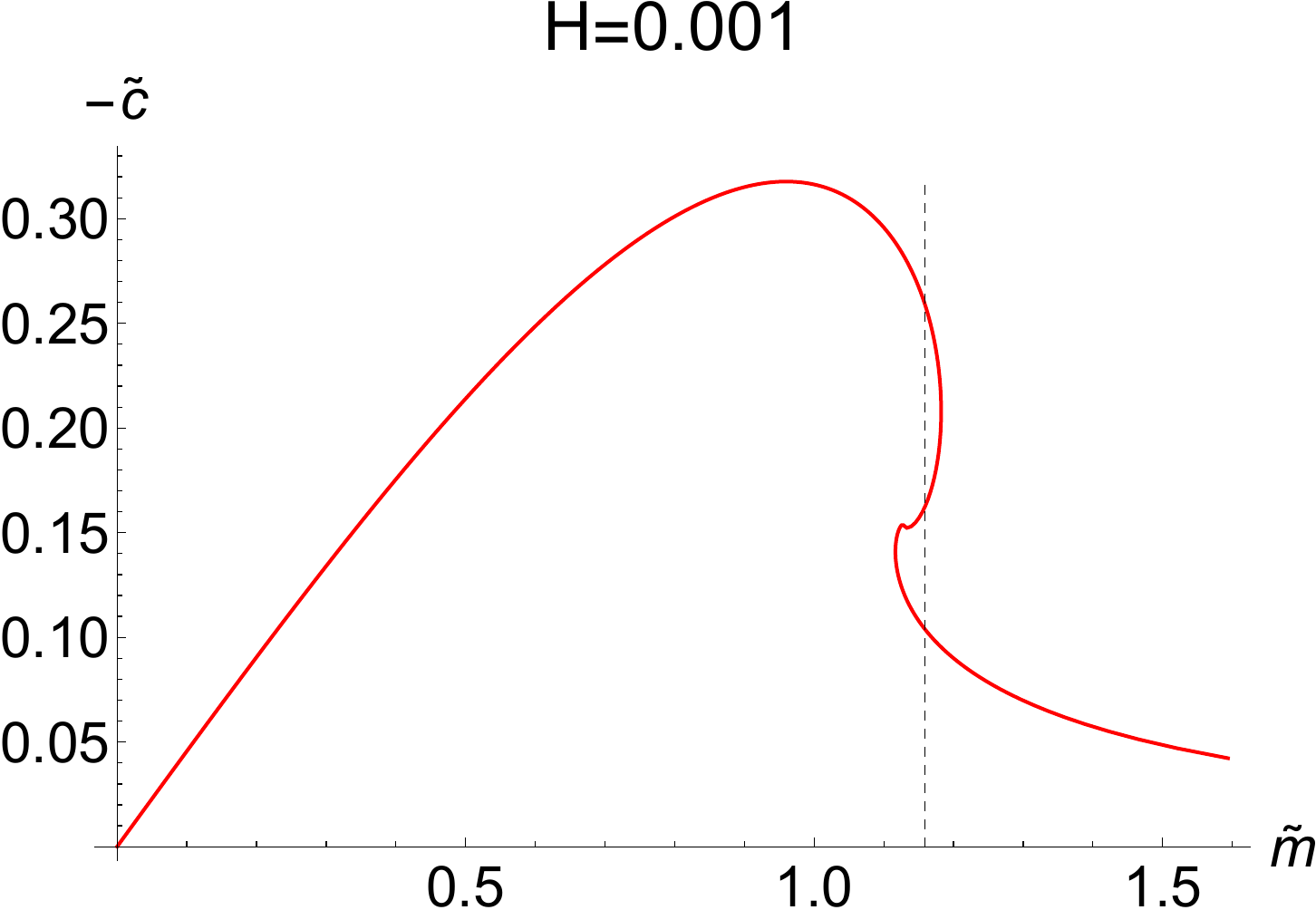}
   \includegraphics[width=6cm]{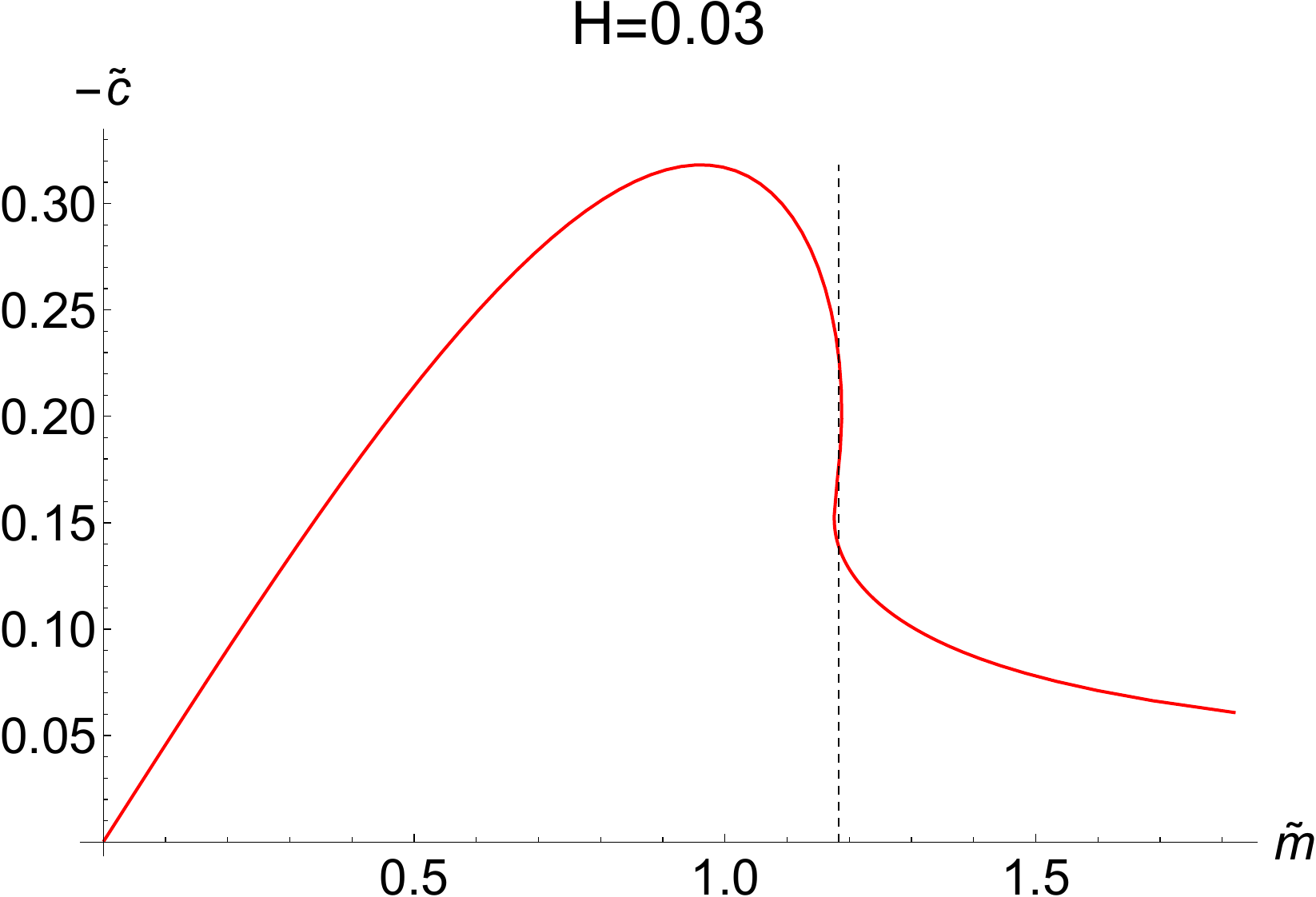}
   \includegraphics[width=6cm]{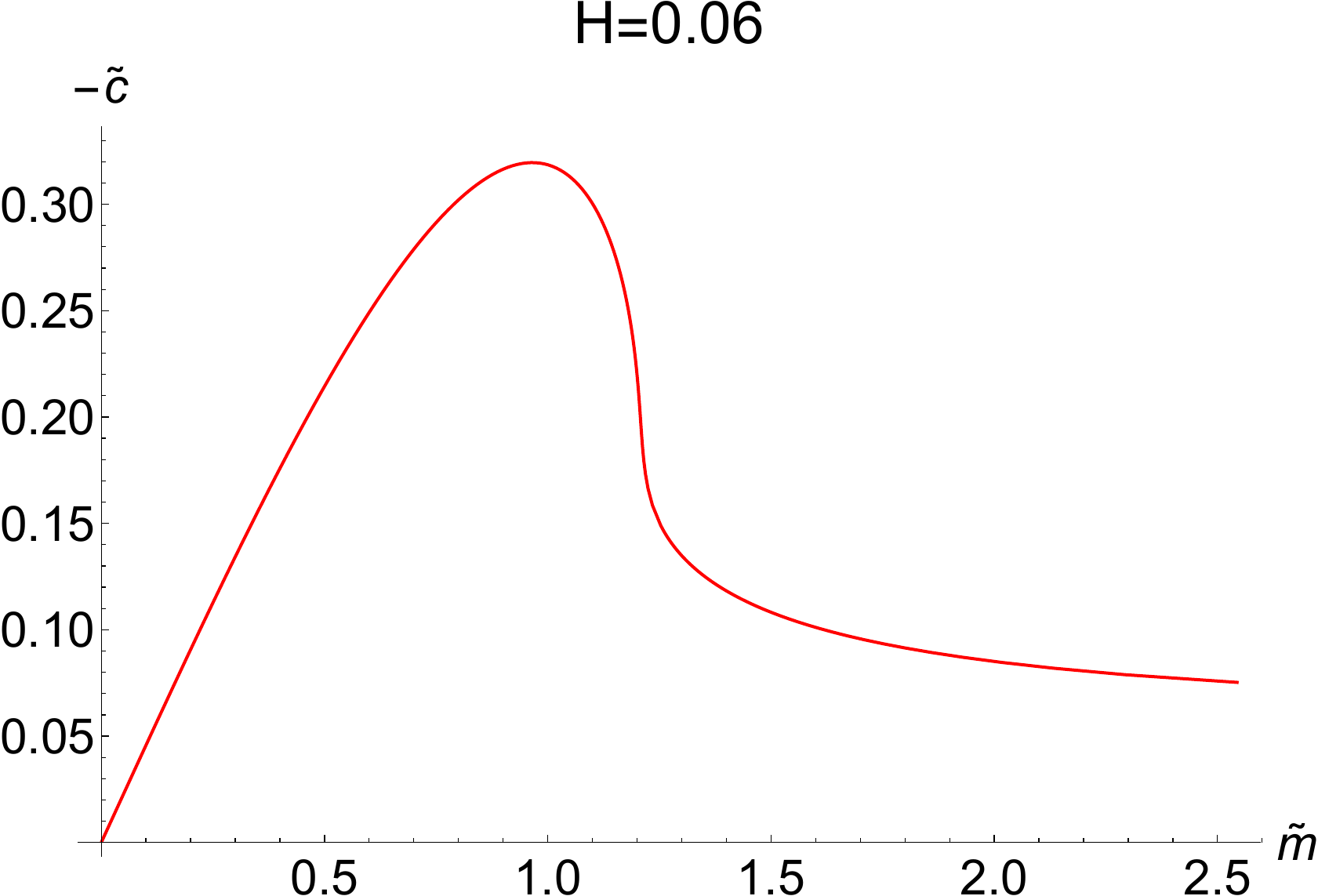}
   \caption{\small Plots of $-\tilde c$ as a function of the bare mass parameter $\tilde m$. For small $H$ there is a clear first order phase transition pattern, which is less pronounced as $H$ and for larger $H$ is replaced by a crossover. The dashed vertical line represent the critical parameter $\tilde m_{cr}$ at which the phase transition takes place.}
   \label{fig:CvsM_all}
\end{figure}

In figure~\ref{fig:CvsM_all} we have presented our results for various values of the parameter $H$. As expected when $H=0$ we recover the first order phase transition pattern reported in ref.~\cite{Myers:2006qr}. Note that the use of  different colours for the equation of state curve represents the fact that the phase transition is a topology change transition between Minkowski (blue) and black hole (red) embeddings (corresponding to a confinement/deconfinement phase transition in the dual gauge theory). 

Next we consider a small amount of $C_{\Delta x_3}$ condensate ($H=0.001$) as one can see the phase transition pattern is extremely close to the one at vanishing $C_{\Delta x_3}$ condensate ($H=0$) but the single colour represents the fact that black hole embeddings are the only stable embeddings and the phase transition takes place entirely within the deconfined phase of the theory. 

In the next plot in Figure~\ref{fig:CvsM_all} we present the equation of state curve for a larger $C_{\Delta x_3}$ condensate ($H=0.03$). As one can see from the plot the first order pattern is much less pronounced, that is: the multivalued region of the $C_m$ versus $m_q$ curve (or $-\tilde c$~versus $\tilde m$) is very narrow. Note that the dashed vertical line in the first three plots represent the critical parameter $\tilde m_{cr}$ at which the phase transition takes place. 

\begin{figure}[h]
   \centering
   \includegraphics[width=6cm]{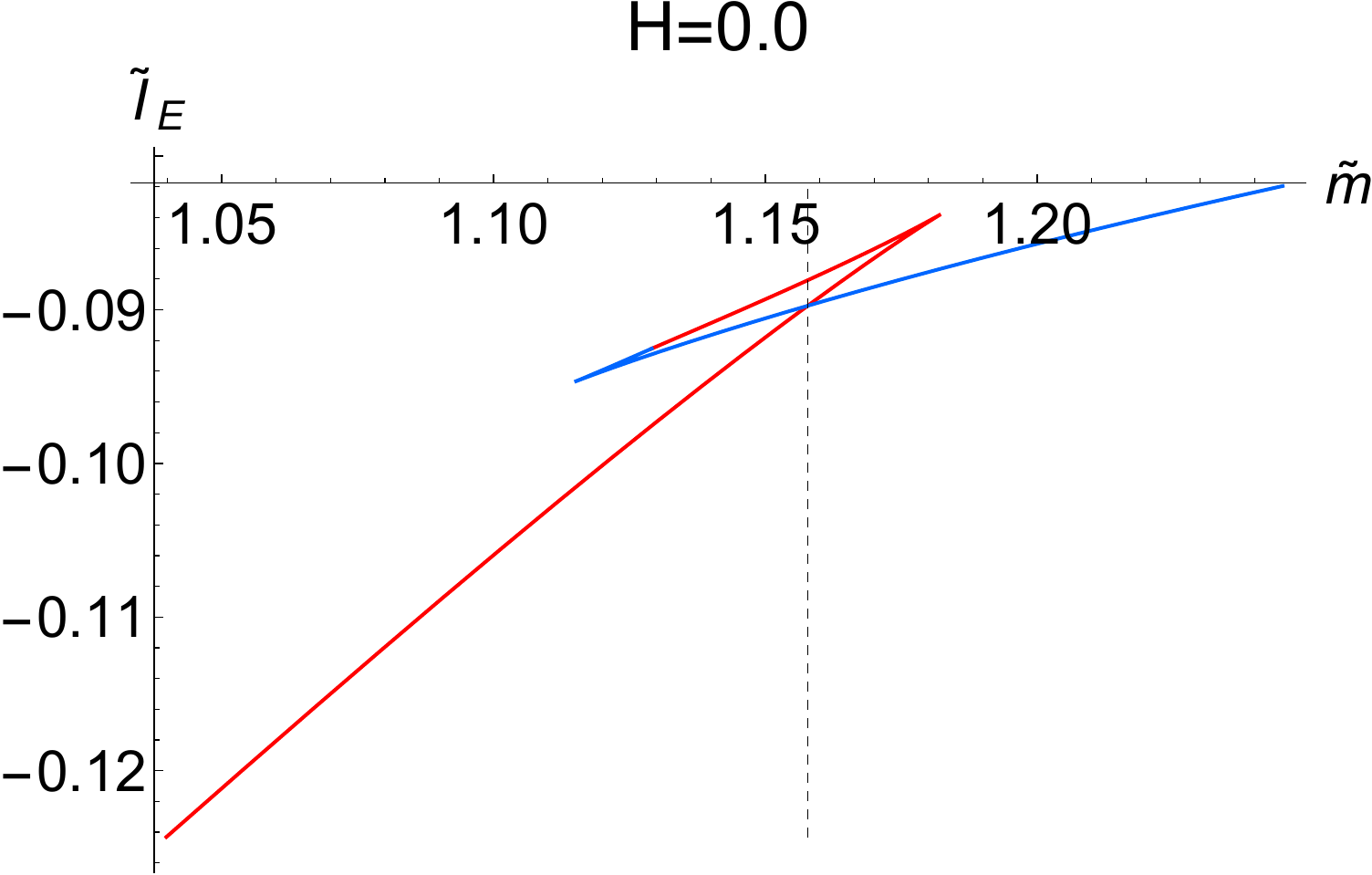}
   \includegraphics[width=6cm]{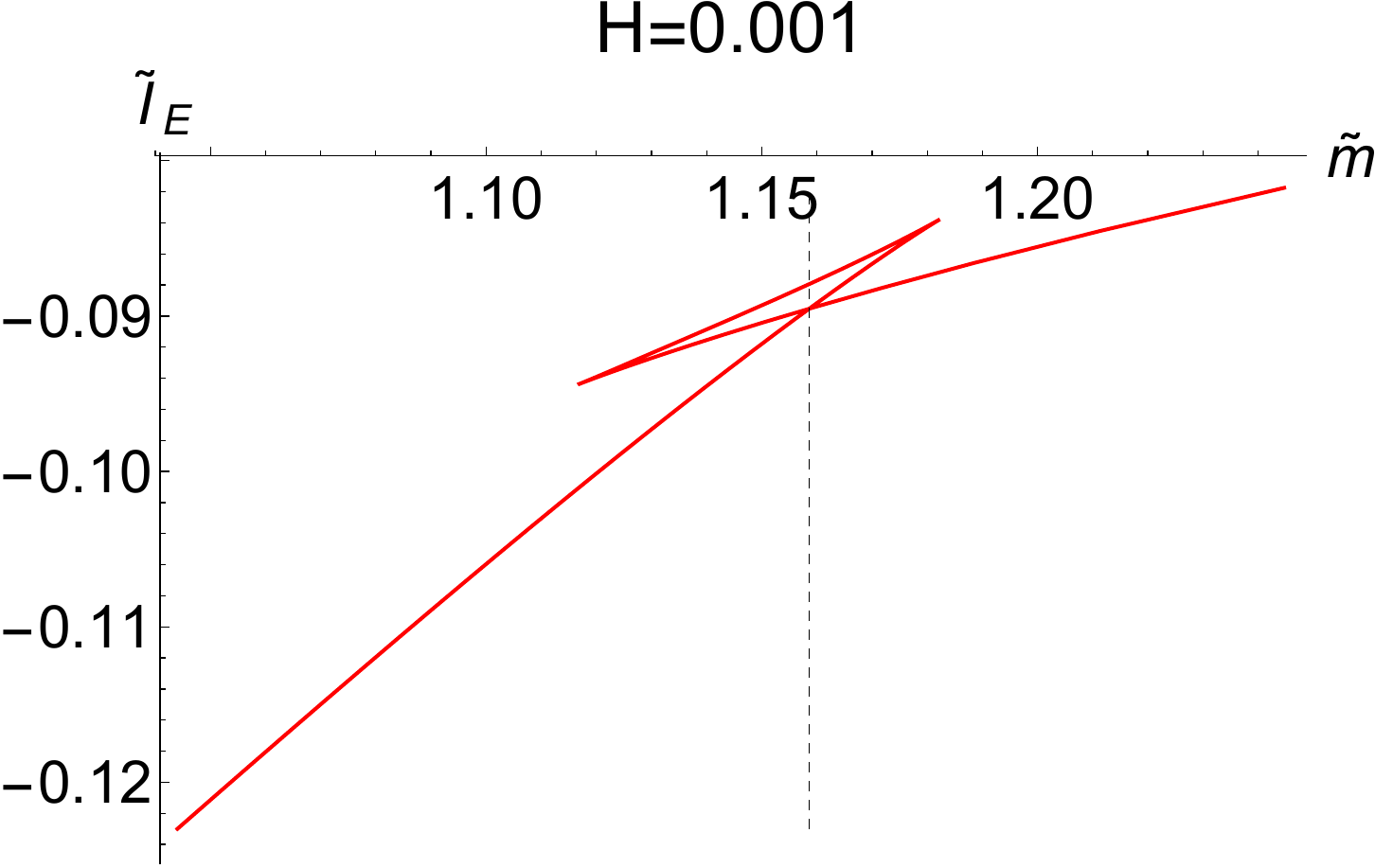}
   \includegraphics[width=6cm]{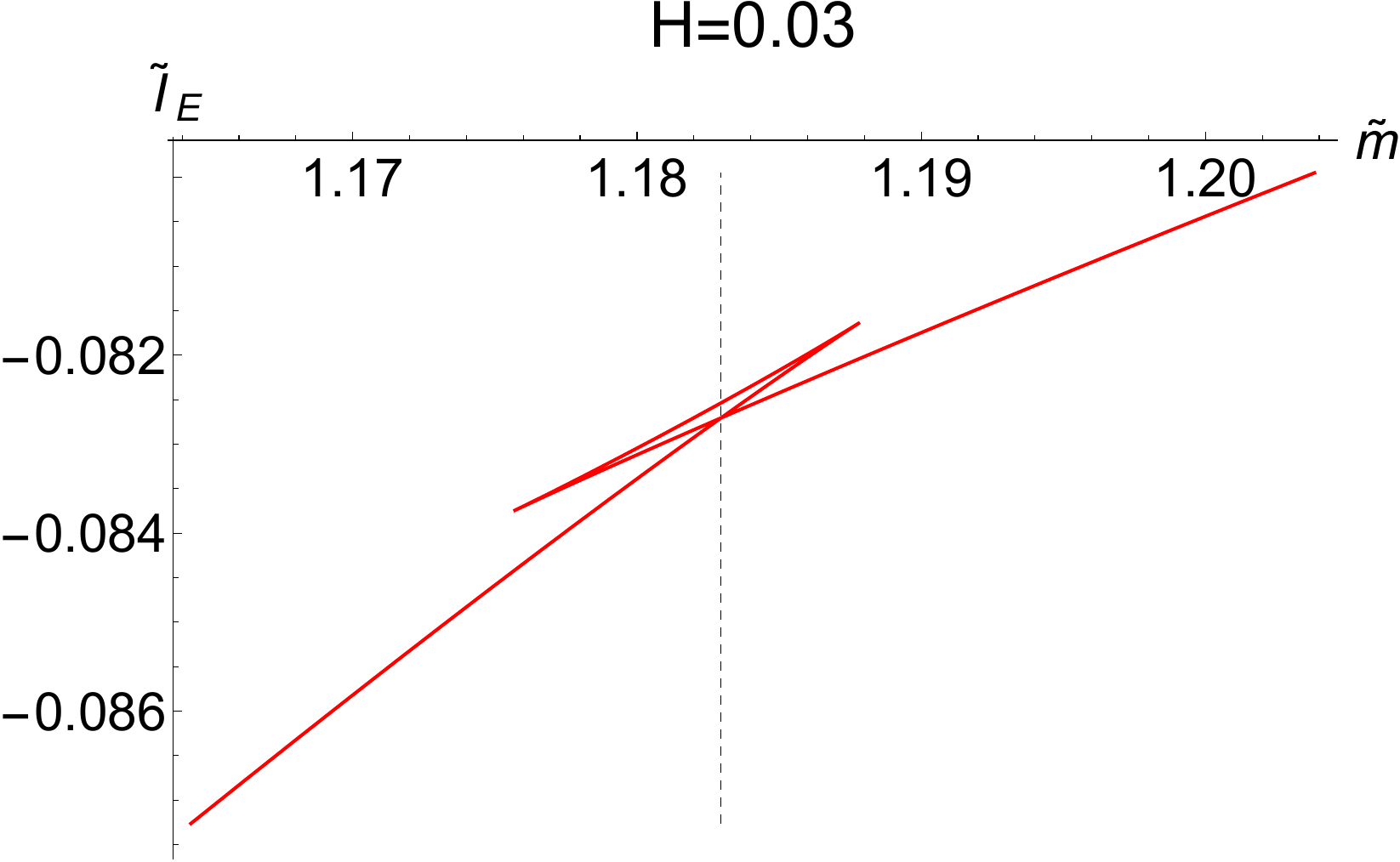}
   \includegraphics[width=6cm]{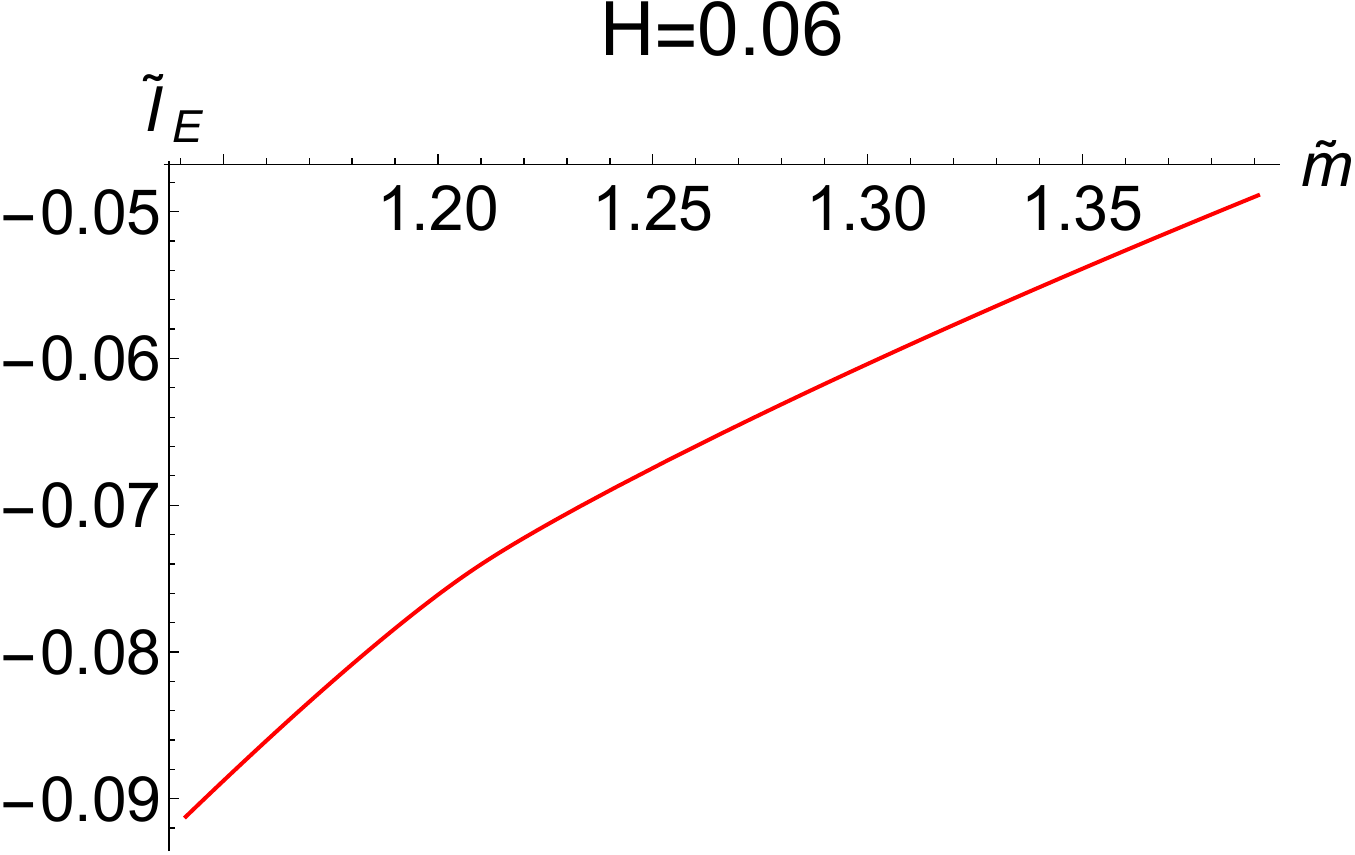}
   \caption{\small Plots of $\tilde I_E$ as a function of the bare mass parameter $\tilde m$. For small $H$ there is a clear first order phase transition pattern, which is less pronounced as $H$ and for larger $H$ is replaced by a crossover. }
   \label{fig:FvsM_all}
\end{figure}

Finally, when we go to even larger values of the $C_{\Delta x_3}$ condensate ($H=0.06$), the first order phase transition disappears as evident in figure~\ref{fig:CvsM_all}. In figure \ref{fig:FvsM_all} we present the corresponding plots of the quantity $\tilde I_E$ defined in equation (\ref{leg-action}) and related to the free energy $\tilde F$ through equation (\ref{tildeF}). One can see that at $H=0.06$ the first order phase transition is replaced by a crossover. 

The observed behaviour suggests that at some $H_{cr}$ the first order phase transition ends on a critical point of a second order phase transition. In the next section we study numerically this critical point with a focus on the critical exponents of the fundamental condensate. 
\subsection{Critical Point}
As the analysis of the previous subsection suggests, for some $H_{cr}$ (in the range $0.03 < H_{cr} < 0.06$) the first order phase transition ends on a critical point of a first order phase transition. Beyond this critical point we have a crossover and the the condensate versus bare mass curve is single valued. It is straightforward to determine numerically the value of $H_{cr}$ at which the multivaluedness of the condensate curve disappears, we obtained $H_{cr} \approx 0.0443835$. 

\begin{figure}[t]
   \centering
   \includegraphics[width=7cm]{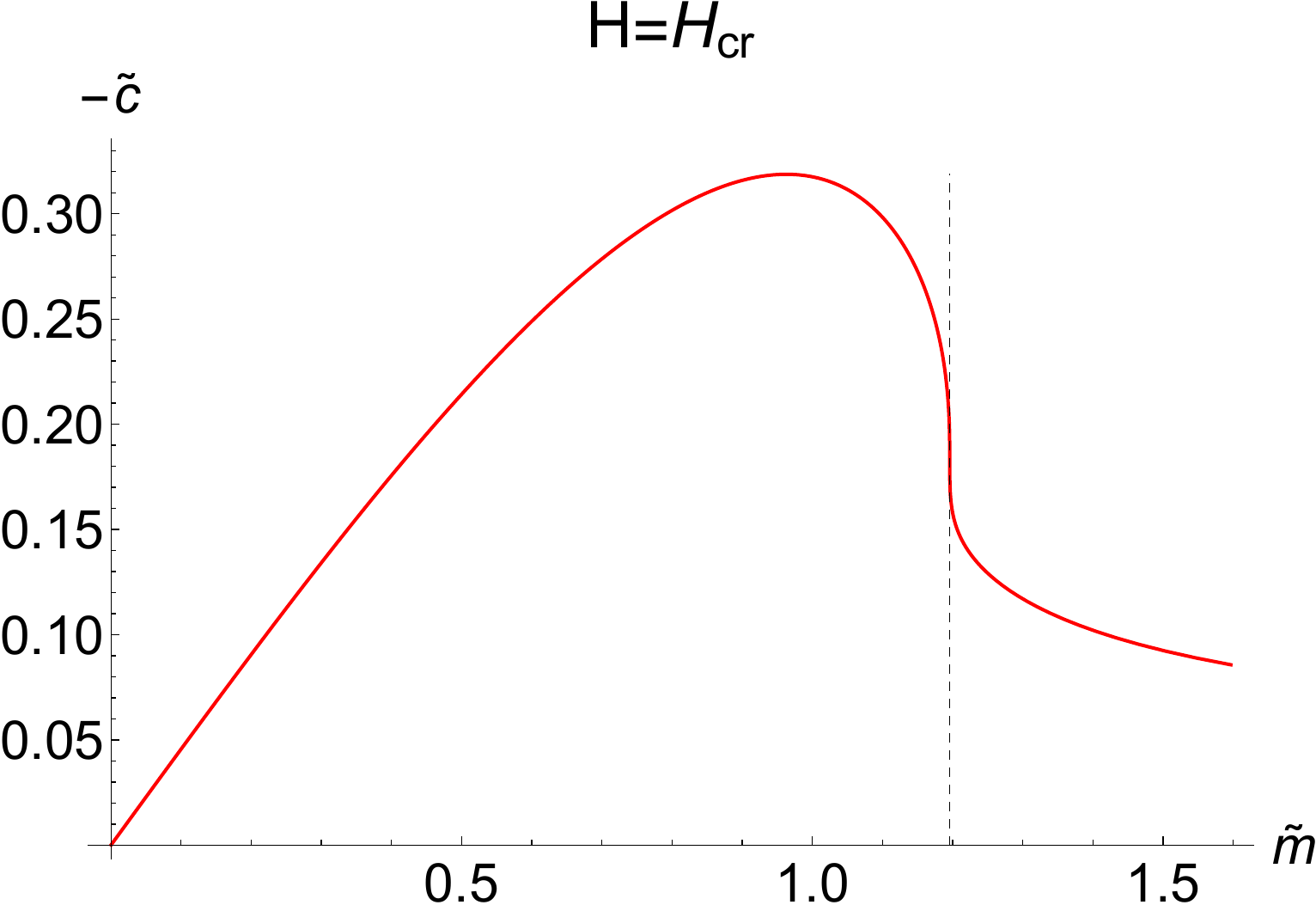}
   \includegraphics[width=7cm]{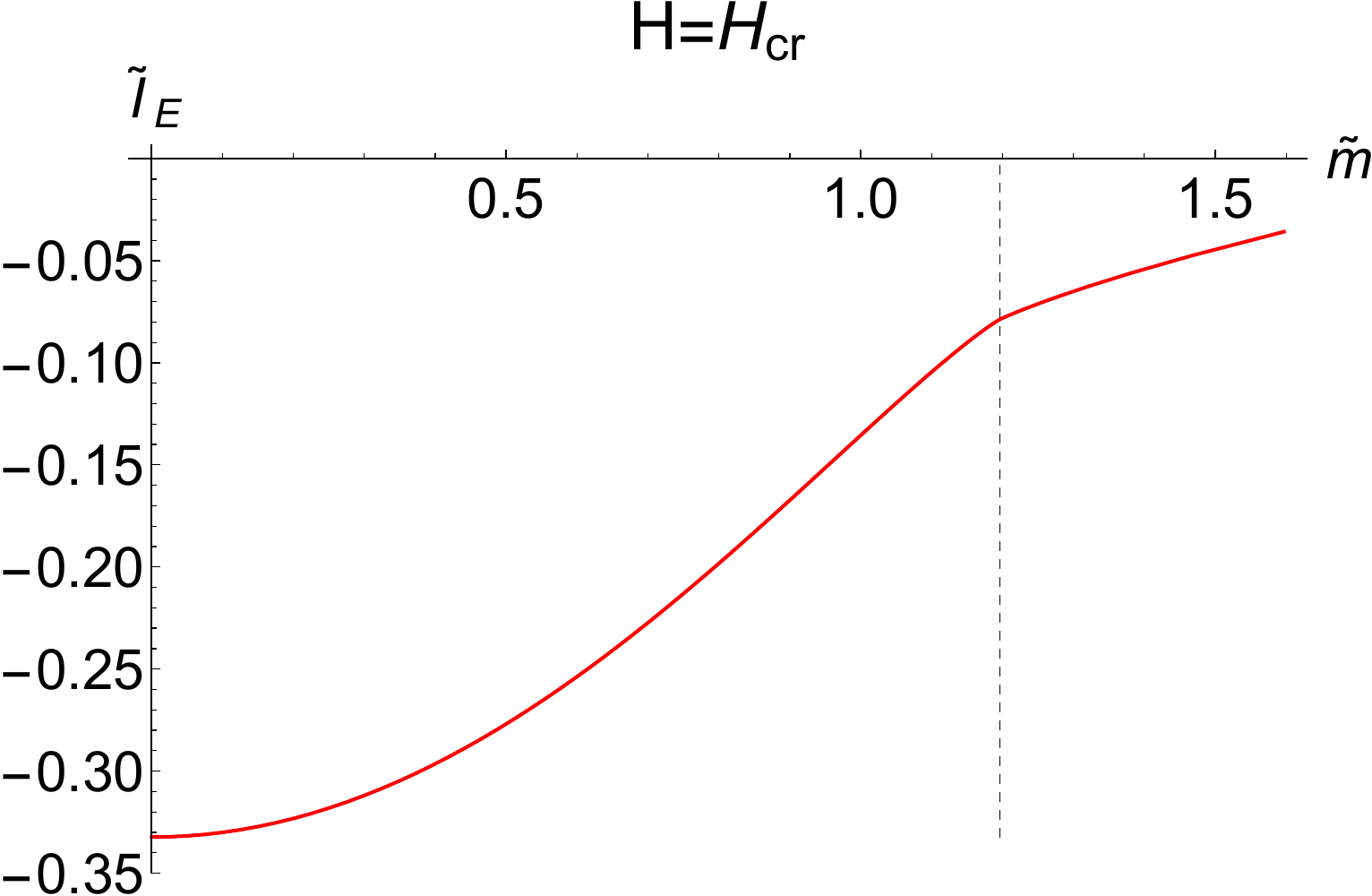}
   \caption{Plots of $-\tilde c$ and $\tilde I_E$ versus $\tilde m$ for $H=H_{cr}$. The Dashed vertical line represents the critical parameter $\tilde m_{cr} = 1.19591$. }
   \label{fig:CriticalCandF}
\end{figure}

\begin{figure}[t]
   \centering
   \begin{subfigure}[]
   \centering
   \includegraphics[width=6.5cm]{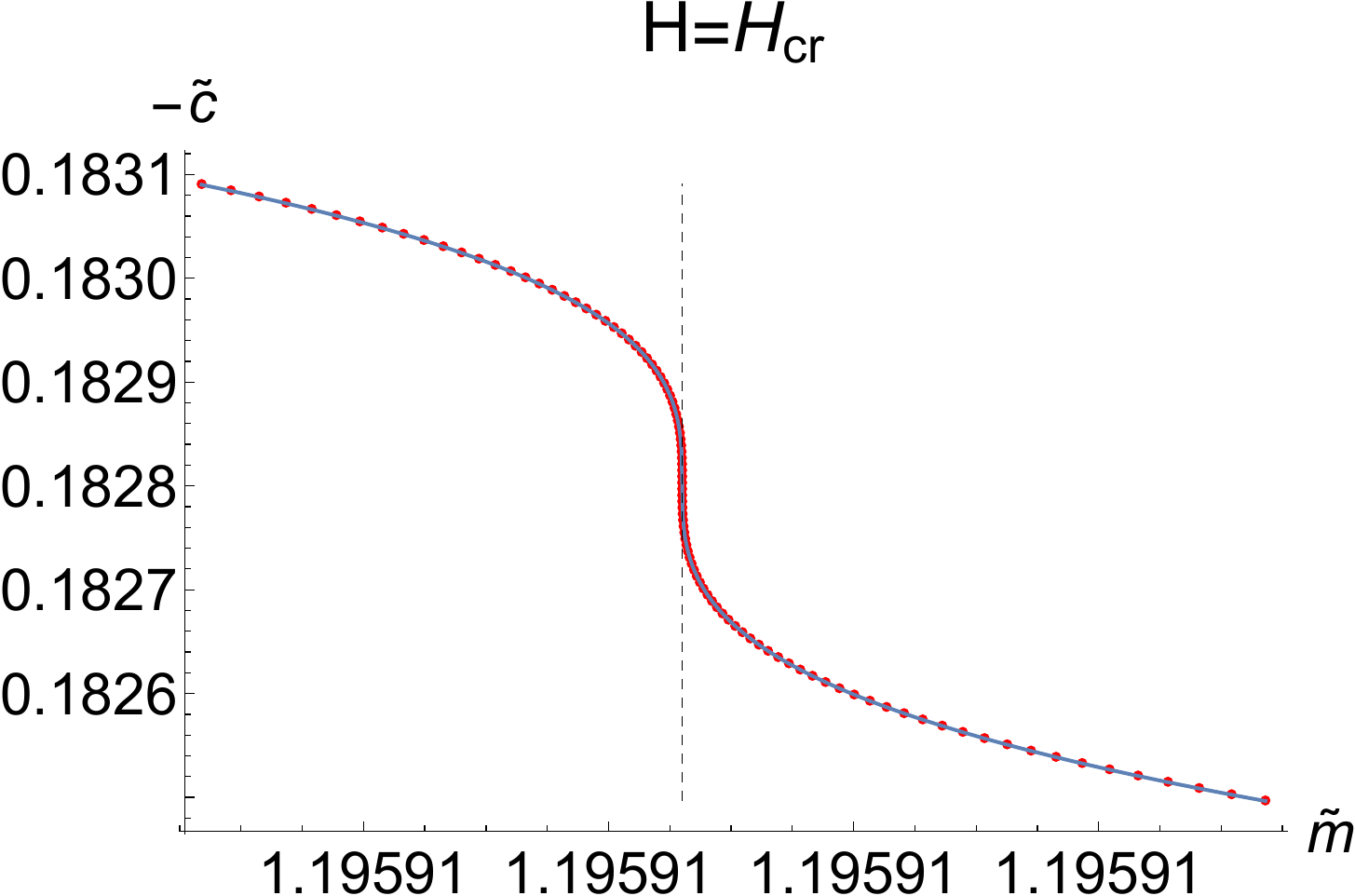}
   \end{subfigure}
   \begin{subfigure}[]
   \centering
   \includegraphics[width=6.5cm]{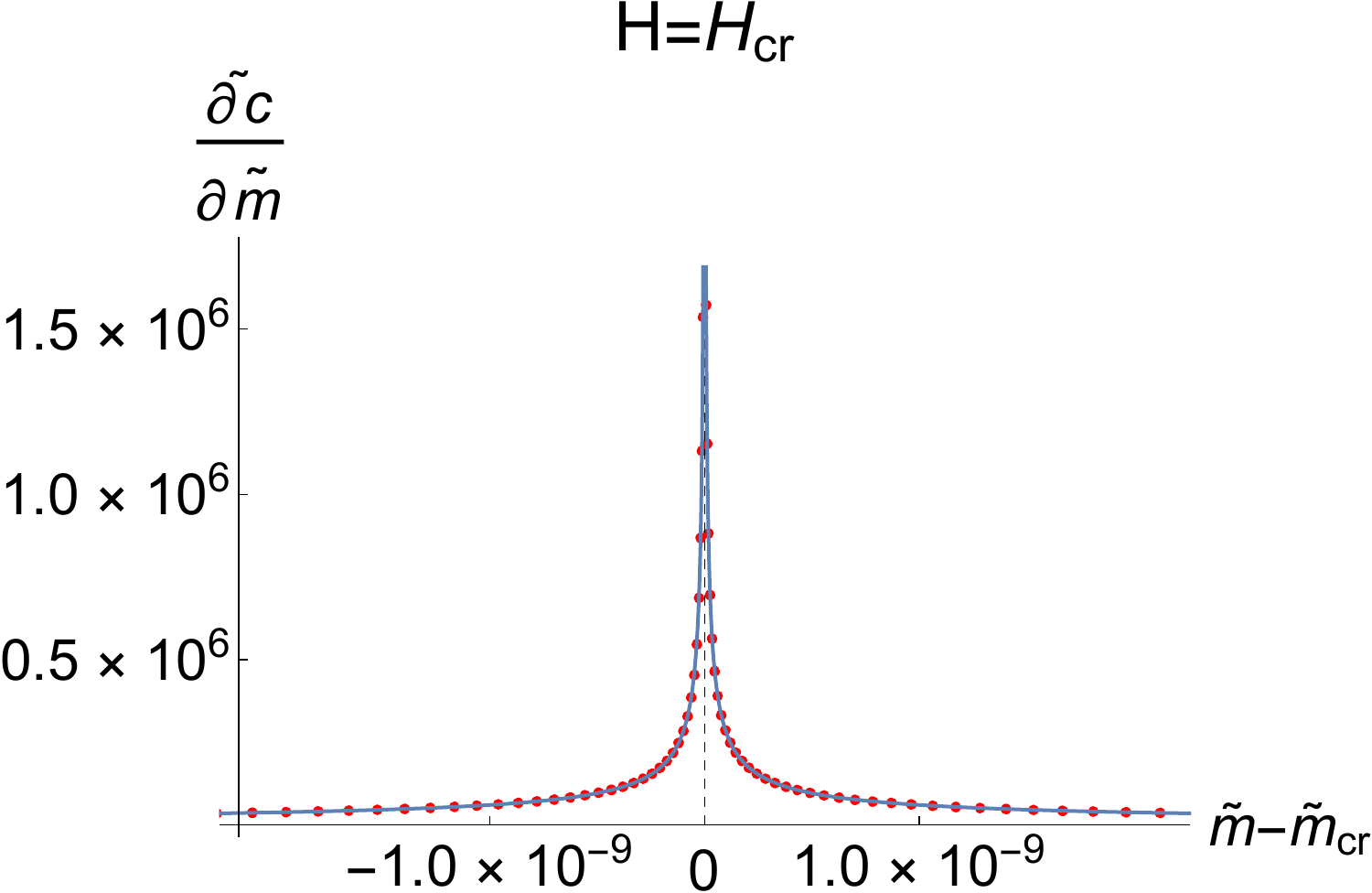}
   \end{subfigure}
   \begin{subfigure}[]
   \centering
   \includegraphics[width=6.5cm]{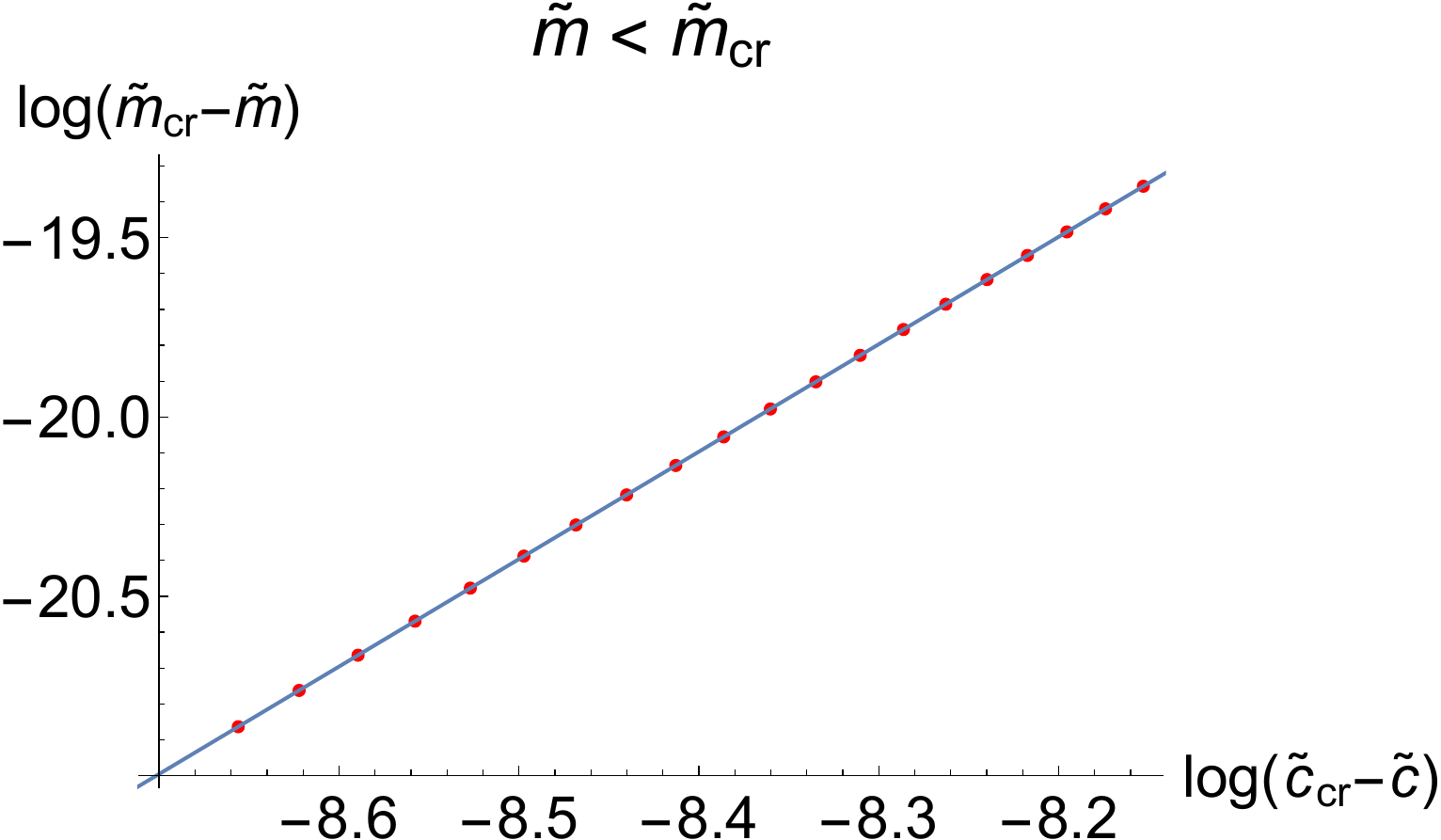}
   \end{subfigure}
   \begin{subfigure}[]
   \centering
   \includegraphics[width=6.5cm]{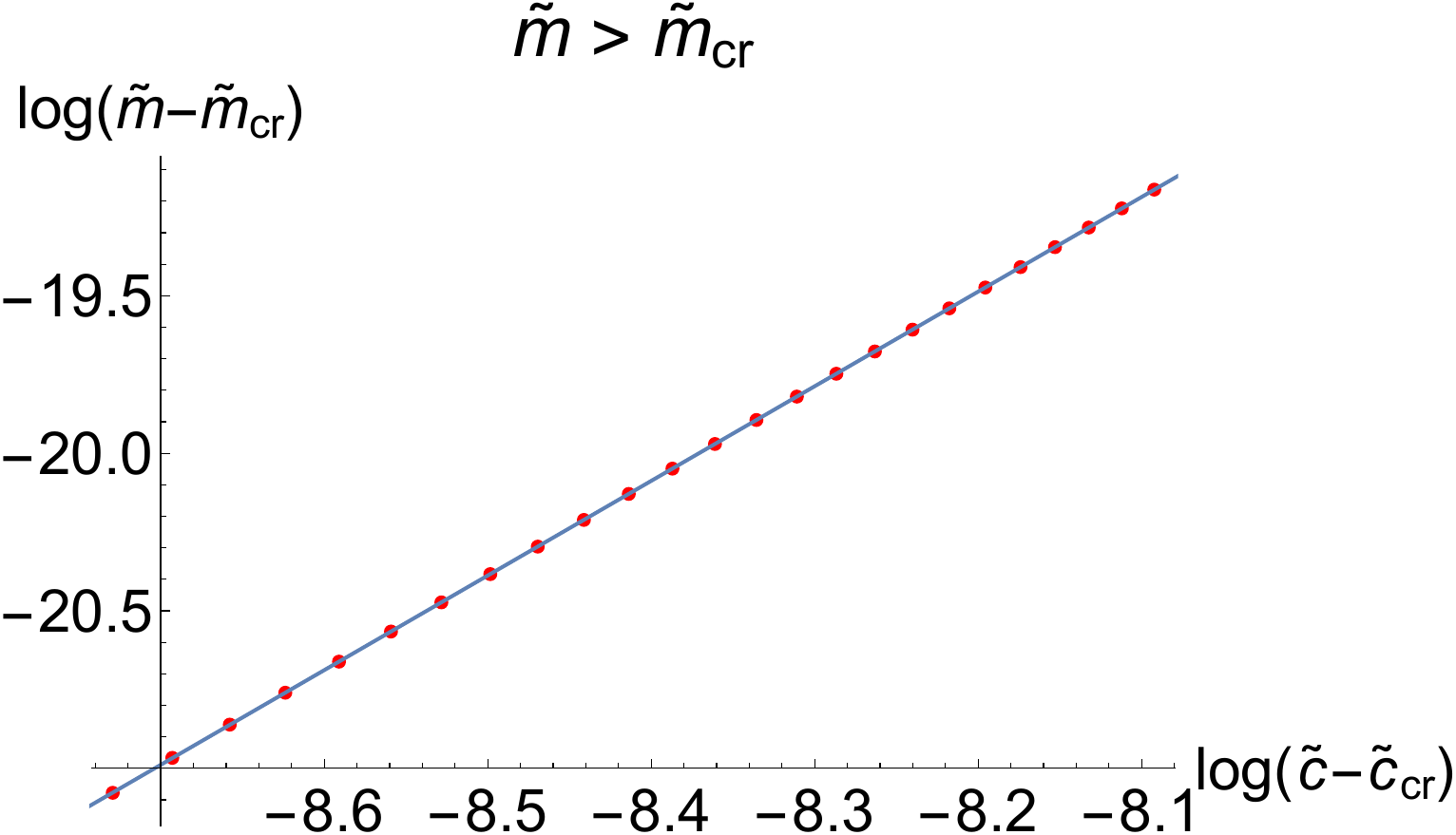}
   \end{subfigure}
   \caption{(a) A plot of $-\tilde c$ versus $\tilde m$ zoomed in near $\tilde m_{cr}$, $\tilde c_{cr}$. (b) A plot of the susceptibility $\tilde c /\tilde m$ versus $\tilde m - \tilde m_{cr}$ near the critical region. (c) A plot of $\log(\tilde m_{cr}-\tilde m)$ versus $\log(\tilde c_{cr}-\tilde c)$ for $\tilde m < \tilde m_{cr}$. (d) A plot of $\log(\tilde m-\tilde m_{cr})$ versus $\log(\tilde c-\tilde c_{cr})$ for $\tilde m > \tilde m_{cr}$.}
   \label{fig:CriticalZoomed}
\end{figure}

In figure \ref{fig:CriticalCandF} we have presented our results for the condensate $C_m$ and the free energy $\tilde F$ as a function of the bare mass. The dashed vertical curve represents the critical parameter $\tilde m_{cr}$ at which the second order phase transition takes place. If we denote by $\tilde c_{cr}$ the critical value $\tilde c_{cr} = \tilde c(\tilde m_{cr})$, and zoom in near the phase transition, we have the relation:
\begin{eqnarray}\label{c_crit}
|\tilde c - \tilde c_{cr}| &\propto & |\tilde m - \tilde m_{cr}|^{\Delta + 1} \ , \\
\frac{\partial^2 \tilde F}{\partial m^2} \propto\frac{\partial \tilde c}{\partial\tilde m} &\propto & |\tilde m - \tilde m_{cr} |^\Delta\label{susc._crit}
\end{eqnarray}
where $\frac{\partial^2 \tilde F}{\partial m^2}$ is proportional is the mass susceptibility of the condensate $C_m$ and $\Delta$ is the corresponding critical exponent. We are going to show numerically that $\Delta = -2/3$ with a very high confidence.   
In Figure \ref{fig:CriticalZoomed} (a) and (b) we have presented a zoom in of the condensate and susceptibility near the the critical region. The dashed vertical line represents the critical parameter $\tilde m_{cr} = 1.19591$, while the continuous curves represent fits with $\Delta = -2/3$ in equations (\ref{c_crit}) and (\ref{susc._crit}). As one can see the fits are excellent. Next rather than feeding in the value of $\Delta$ we extract its value using by taking a logarithm on both sides in equation (\ref{c_crit}) and using a linear regression. In Figure \ref{fig:CriticalZoomed} (c) and (d) we have plotted $\log |\tilde m -\tilde m_{cr}|$ versus  $\log |\tilde c -\tilde c_{cr}|$ for both $\tilde m < \tilde m_{cr}$ and $\tilde m > \tilde m_{cr}$. By taking the mean of the two slopes we can estimate $\Delta$ and its standard deviation. Our results is:
\begin{equation}
\Delta = -0.66666 \pm 0.00002\ ,
\end{equation}
which is extremely close to $\Delta = -2/3$. Interestingly, this is the same critical exponent as the one reported in ref.~\cite{Filev:2014mwa}\footnote{The quantity studied ref.~\cite{Filev:2014mwa} was $\gamma = \Delta+1 = 1/3$.}, where the same holographic technique (introducing a non-zero $U(1)$ flux through a two sphere) was considered to deform the first order phase transition into a second order one. 

\section{Discussion}

In this paper we consider the holographic gauge theory dual to the D3/D5-brane intersection with a $U(1)$ flux on the transverse two-sphere wrapped by the D5--brane probes. We consider both the finite temperature and zero temperature phases of the theory.

In the zero temperature case Ramond-Ramond charge conservation causes the D5--brane embeddings to bend along the Neumann-Dirichlet direction ($x_3$) carrying the charge to infinity. Remarkably, these solutions preserve the ${\cal N}=2$ supersymmetry of the D3/D5 intersection  and interpolate between D5--brane embeddings at large radial distance and D3--brane probes (parallel to the stack of D3--branes sourcing the geometry). This not only solves the charge conservation problem but splits the 1+3 dimensional Super Yang-Mills theory into regions with different ranks of the gauge group. That is the 1+2 dimensional defect (introduced by the D5--branes) behaves as a domain wall. In the case of zero bare mass the theories on both sides of the domain wall are \cite{Myers:2008me} $SU(N_c +k)$ and $SU(N_c)$, while in the case of finite bare mass we showed that the groups are $SU(N_c)\times SU(k)$ and $SU(N_c)$. Given that the field content of the ${\cal N}=2$ theory is completely fixed by supersymmetry we extended the field theory interpretation of ref.~\cite{Myers:2008me} to a non-zero bare mass.

At finite temperature, a D3--brane probe is no longer ``natural'' in the sense that D3--brane probes parallel to the field theory directions are no longer stable solutions. As a result a solution interpolating between a D5--brane and a (parallel) D3--brane embedding is no longer possible. To avoid violation of Ramond-Ramond charge conservation the Mikowski embeddings develop a D3--brane throat connecting them to the black hole horizon, while still developing a profile in the Neumann-Dirichlet direction $x_3$. Note that even though Black hole embeddings do not extend all the way to $x_3 = -\infty$ the continuity of the horizon (the horizon is extended along $x_3$) requires the Ramond-Ramond charge to reach infinity \cite{Myers:2008me} and the domain wall interpretation is valid at finite temperature too. Considering equation (\ref{profileCFT}) it is natural to speculate that at finite bare mass the large gauge group is $SU(N_c)\times SU(k)$ and only at vanishing bare mass the group is promoted to $SU(N_c + K)$.

Even though Mikowski embeddings violate charge conservation, they can still be approximately realised within the class of black hole embeddings and the first order phase transition between the two classes of embeddings is no-longer a topology change transition.\footnote{This is analogous to the D3/D7 system in the presence of a baryon density studied in ref.~\cite{Kobayashi:2006sb}.} The latter opens up the possibility to deform the first order phase transition to a second order one. Furthermore, the non-trivial profile along $x_3$ is reflected in a non-zero condensate $C_{\Delta x_3}$ thermodynamically conjugated to the position of the defect and proportional to the $U(1)$ flux on the transverse two-sphere. Increasing the flux (the $C_{\Delta x_3}$ condensate) lowers the latent heat of the first order phase transition and for sufficiently strong flux it ends on a critical point of a second order phase transition. Our analysis of the critical point showed that the mass susceptibility diverges at criticality with a critical exponent of $\Delta = -2/3$. Interestingly, this is the same critical exponent as the one reported in ref.~\cite{Filev:2014mwa}, where the holographic set-up also includes a constant $U(1)$ flux through a shrinking $S^2$ cycle. 

Given that the dual field theory is 2 + 1 dimensional the critical point that we studied might be relevant to a condensed matter system exhibiting critical behaviour with the same critical exponents. One might be able to use the holographic set-up as a toy model to study the near critical regime of such condensed matter system. 

Our work can be extended in several possible directions. The most straightforward extension is to supplement our thermodynamical analysis with a study of the quantum fluctuations/meson spectra of the theory. Another extension is considering the holographic set-up in the presence of an external electric field or R-charge chemical potential, which will enable us to study quantum phase transitions. Finally, it would be interesting to understand better the sumersymmetric phase of the theory and the observed domain wall.

\section*{Acknowledgments}
We would like to thank Denjoe O'Connor and Nikolay Bobev for useful comments and suggestions. The work of VF was supported in part by Bulgarian NSF grants DN08/3 and H28/5. The work of RR was supported in part by Bulgarian NSF grant H28/5.

\end{document}